\documentclass[twocolumn]{aastex63}

\defcitealias{Mikolaitis2018}{M18}
\defcitealias{Mikolaitis2019}{M19}

\newcommand{\kms}{\,km\,s$^{-1}$\,}
\newcommand{\Li}{\,$^{7}$Li\,}
\newcommand{\Teff}{\,$T_{\rm{eff}}$\,}
\newcommand{\logg}{\,log\,${{g}}$\,}
\newcommand{\FeH}{\,[Fe/H]\,}
\newcommand{\ALi}{\,$A$({Li})\,}
\newcommand{\MS}{\,$M_{\odot}$\,}
\newcommand{\vt}{\,$v_{\rm t}$\,}
\newcommand{\Vsini}{\,$v$\,sin\,$i$}

\renewcommand{\thempfootnote}{\roman{mpfootnote}}
\defcitealias{Mikolaitis18}{Paper~I}
\defcitealias{Mikolaitis19}{Paper~II}

\received{December 10, 2019}
\revised{January 3, 2020}
\accepted{January 4, 2020}

\submitjournal{AJ}

\shorttitle{Lithium, carbon, and oxygen abundances}
\shortauthors{Stonkut\.{e} et al.}

\begin{document}

\title{
High-resolution Spectroscopic Study of Dwarf Stars in the Northern Sky: Lithium, Carbon, and Oxygen Abundances}

\correspondingauthor{E. Stonkut\.{e}}
\email{edita.stonkute@tfai.vu.lt}

\author{E. Stonkut\.{e}}
\affiliation{Astronomical Observatory, Institute of Theoretical Physics and Astronomy, 
Vilnius University, \\ Sauletekio av. 3, 10257 Vilnius, Lithuania}

\author{Y. Chorniy}
\affiliation{Astronomical Observatory, Institute of Theoretical Physics and Astronomy, 
Vilnius University, \\ Sauletekio av. 3, 10257 Vilnius, Lithuania}

\author{G. Tautvai\v{s}ien\.{e}}
\affiliation{Astronomical Observatory, Institute of Theoretical Physics and Astronomy, 
Vilnius University, \\ Sauletekio av. 3, 10257 Vilnius, Lithuania}

\author{A. Drazdauskas}
\affiliation{Astronomical Observatory, Institute of Theoretical Physics and Astronomy, 
Vilnius University, \\ Sauletekio av. 3, 10257 Vilnius, Lithuania}

\author{R. Minkevi\v{c}i\={u}t\.{e}}
\affiliation{Astronomical Observatory, Institute of Theoretical Physics and Astronomy, 
Vilnius University, \\ Sauletekio av. 3, 10257 Vilnius, Lithuania}

\author{\v{S}. Mikolaitis} 
\affiliation{Astronomical Observatory, Institute of Theoretical Physics and Astronomy, 
Vilnius University, \\ Sauletekio av. 3, 10257 Vilnius, Lithuania}

\author{H. Kjeldsen}
\affiliation{Astronomical Observatory, Institute of Theoretical Physics and Astronomy, 
Vilnius University, \\ Sauletekio av. 3, 10257 Vilnius, Lithuania}
\affiliation{Stellar Astrophysics Centre, Department of Physics and Astronomy, Aarhus University, \\ Ny Munkegade 120, DK-8000 Aarhus C, Denmark}

\author{C. von Essen}
\affiliation{Astronomical Observatory, Institute of Theoretical Physics and Astronomy, 
Vilnius University, \\ Sauletekio av. 3, 10257 Vilnius, Lithuania}
\affiliation{Stellar Astrophysics Centre, Department of Physics and Astronomy, Aarhus University, \\ Ny Munkegade 120, DK-8000 Aarhus C, Denmark}

\author{E. Pak\v{s}tien\.{e}}
\affiliation{Astronomical Observatory, Institute of Theoretical Physics and Astronomy, 
Vilnius University, \\ Sauletekio av. 3, 10257 Vilnius, Lithuania}

\author{V. Bagdonas}
\affiliation{Astronomical Observatory, Institute of Theoretical Physics and Astronomy, 
Vilnius University, \\ Sauletekio av. 3, 10257 Vilnius, Lithuania}




\begin{abstract}

Abundances of lithium, carbon, and oxygen have been derived using spectral synthesis for a sample of 249 bright F, G, and K Northern Hemisphere dwarf stars from the high-resolution spectra acquired with the VUES spectrograph at the Mol{\. e}tai Astronomical Observatory of Vilnius University. 
The sample stars have metallicities, effective temperatures, and ages between \mbox{(--0.7 $\div$ 0.4)~dex}, \mbox{(5000 $\div$ 6900)~K}, \mbox{(1 $\div$ 12)~Gyr}, accordingly.
We confirm a so far unexplained lithium abundance decrease at supersolar metallicities -- $A$(Li) in our sample stars, which  drop by 0.7~dex in the [Fe/H] range from +0.10~to~+0.55~dex. Furthermore, we identified stars with similar ages, atmospheric parameters, and rotational velocities, but with significantly different lithium abundances, which suggests that additional specific evolutionary factors should be taken into account while interpreting the stellar lithium content. 
Nine stars with predominantly supersolar metallicities, i.e. about 12\,\% among 78 stars with C and O abundances determined, have the C/O number ratios larger than 0.65, thus may form carbon-rich rocky planets. Ten planet-hosting stars, available in our sample, do not show a discernible difference from the stars with no planets detected regarding their lithium content. 

\end{abstract}



\section{Introduction} \label{sec:Introduction}


In this work, we focus on a homogeneous abundance determination of the three important chemical elements -- lithium, carbon, and oxygen -- in a sample of dwarf stars and analyze the results in the light of extrasolar planet search surveys (e.g., NASA's Transiting Exoplanet Survey Satellite ({\sc {TESS}}; \citealt{Ricker15}) and the upcoming  ESA's PLAnetary Transits and Oscillations of stars ({\sc{PLATO}}; \citealt{Rauer14}) which will focus on the bright targets, of which only a third has any spectroscopic studies.

The complex pattern of lithium abundance -- \ALi, observed in different types of stars in the Milky Way is a subject of ongoing discussions. \Li~has three different nucleosynthesis sites -- Galactic cosmic rays, stars, and primordial nucleosynthesis.
Lithium is one of four elements synthesized in the primordial nucleosynthesis with the initial value of \ALi$ = 2.72$~dex. This value is predicted from the Big Bang nucleosynthesis models based on the results from the \textit{Wilkinson Microwave Anisotropy Probe} \citep{Cyburt08} and is approximately three to four times higher than the observationally determined value from halo dwarfs \citep{Spite82}. Stars are producing the majority of \Li~in the Galaxy. It is still unclear whether the sources are red giants, asymptotic giant branch stars, novae, or core-collapse supernova. 
Carbon and oxygen, on the other hand, were synthesized in the post Big Bang epoch. Both elements, however, were formed by different processes. Carbon acts as a catalyst in the CNO-cycle converting H to He. Carbon production occurs in stellar interiors, then it is dredged up from cores and mainly released into the interstellar medium by massive stellar winds, driven by radiation pressure from high-mass stars \citep[e.g.,][]{Gustafsson99}. The low- and intermediate-mass stars contribute as well, but their relative importance and yields are not well understood. Carbon is one of the required elements for life, as we know it, and plays an important role in searching for habitable exoplanets.

Oxygen is the third most common element in the universe and is produced by hydrostatic burning in massive stars and then essentially mostly dispersed by the Type\,II supernovae. The oxygen abundance is one of the key tracers of the formation and evolution of planets, stars, and galaxies. The abundances of oxygen in atmospheres of stars, differently from carbon, remain almost constant during the lifetime of stars and thus can be studied to trace back the Galactic chemical evolution.

Every new analysis of these chemical elements in stellar atmospheres contributes to the ongoing discussion of their importance in understanding the Galaxy evolution and whether they constitute a star--planet connection \citep[see, e.g.,][]{Delgado10,Gonzalez10,Brugamyer11,Petigura11,Schuler11,Ramirez12,Brewer16a,Spina16a,Luck17,Adibekyan18,Bensby18,Fu18,Luck18, Nissen18, Guiglion19,Pavlenko19}. 

For example, the sites of lithium production are identified, but their input for the lithium enrichment of the ISM is still debatable \citep{Bensby18, Fu18,Guiglion19}. The poorly understood mechanisms of \Li~depletion and/or synthesis observed in F-, G-, and K-type stars make it difficult to constrain the Galactic chemical evolution models \citep{Cescutti19}. 
Classical evolutionary models predict that the Li abundances in the main sequence (MS) stars should depend uniquely on the stellar main parameters such as the effective temperature, metallicity, age, or chromospheric activity. According to the work by \citet{Israelian04}, the \Teff (or mass) is the main parameter responsible for the Li depletion in solar-type stars. The second depletion driving parameter is the stellar age, while the third parameter could be a metallicity and/or rotation. 
Classical chemical evolution models show a smooth increase of lithium in relation to metallicity \citep{Romano01,Prantzos12}. Those models have some limitations and neglect several key physical processes such as gravitational settling, thermal diffusion, and radiative acceleration, rotational mixing, mass loss -- any of which could be critical for the interpretation of \Li~abundances in solar-type stars \citep{Fu15}. 

The same stars that produce \Li, destroy it as well. Lithium nuclei are burned through the proton capture when they are exposed to temperatures $2.5 \times 10^6$~K. \Li~depletion observed in the Sun is inconsistent with the classical models. In order to explain the observations, lithium must be transported from the convection zone to the hot layers where the temperature is more than $2.5 \times 10^6$~K \citep{Israelian04}. 

From observational data we also see that \Li~abundances in stars above the solar metallicity show lower values than those at the solar metallicity, suggesting a puzzling decrease of lithium. Recent high-resolution spectroscopic surveys of the Milky Way disk have reported that lithium abundances in the solar neighborhood decrease at supersolar metallicities.  \citet{DelgadoMena15} were some of the first to show that the \Li~abundance decreases with ${\rm [Fe/H]} > 0$ and later works by \cite{Guiglion16} in the AMBRE project, \cite{Fu18} in the $Gaia$-ESO Survey, and \cite{Bensby18} from combined data of various spectrographs (e.g., FEROS and MIKE) confirmed this result.
The studies by \cite{Guiglion16} and \cite{Prantzos17} tried to explain the \Li~decreases at supersolar metallicities with stellar migration. The behavior of \Li~observed in high-resolution spectroscopic surveys can be explained by the interplay of mixed populations that originate in the inner regions of the Galaxy disk. The recent work by \cite{Guiglion19} claims that stars have lower \Li~content as a consequence of inside-out disk formation and radial migration. 

Other mechanisms such as the presence of planets have been proposed to be responsible for additional \Li~depletion.  The work of \citet{King97} was the first to suggest a connection between \Li~depletion and planet hosts after finding a difference in \Li~abundance for the very similar stars of the double system 16\,Cyg, in which one of the stars is known to be a host to a Jupiter sized planet.

\cite{Israelian04} and \cite{Chen06} have suggested that Li depletion could be a result of planet migration, creating  instability that produces effective mixing. The protoplanetary disk retains a large amount of angular momentum and therefore creates some rotational breaking in the host stars during the pre-MS phase inducing an increased mixing. \citet{Theado12} suggested that the accretion of metal-rich planetary material onto a star in its early phases could induce the thermohaline convection below the convective zone and lead to the extra \Li~depletion in stars.

The discussion on \Li~depletion on exoplanets hosts was revived by \citet{Figueira14} and \citet{DelgadoMena15}, who concluded that exoplanet hosts show a significant \Li~depletion when compared to hosts without planets. Later work by \citet{Bensby18} found that there is no difference in \Li~abundances in stars with detected planets compared to those with no (at the moment) detected planets.


In the case of the evolution of carbon and oxygen abundances in the Galaxy, they can be used to set constraints on stellar nucleosynthesis and to help understand the formation and evolution of the Milky Way. The information about the origins and evolution of carbon and oxygen may be obtained from differences in the elemental ratios [C/Fe], [O/Fe], and [C/O] when looking at two stellar populations -- thin and thick disk. 
The study by \citet{Reddy06} has found evidence of a systematic difference in [C/Fe] between thin- and thick-disk stars. This difference was not confirmed by \citet{Bensby06}. Furthermore, in the work by \citet{Cescutti09} it was shown that there is a clear distinction between the trends of [C/O] in the thin and thick Galactic disk {comparing the results from observational data with the predictions of Galactic chemical evolution models}. \citet{Cescutti09} suggest that the differences in [C/O] versus [O/H] for the two disks show that the thick disk is not made from the thin-disk material by dynamical heating of the thin disk, or that it was not possible to make the lower metallicity thin disk from the thick-disk material.

The past decade induced the theoretical and observational studies of carbon and oxygen abundances in stars in the context of determining a composition of extrasolar terrestrial planets, assuming that the composition of the host star and its protoplanetary disk is interrelated \citep[see, e.g.,][and references therein]{Bond10, Petigura11,Nissen14, DelgadoMena15, Madhusudhan16,Bedell18}. C and O and their elemental number rations C/O\footnote{C/O is defined as {\rm N$_\mathrm{C}$/N$_\mathrm{O}$}, where {\rm N$_\mathrm{C}$} and {\rm N$_\mathrm{O}$} are the number densities of carbon and oxygen nuclei, respectively. [C/O] is the solar-normalized logarithmic ratio.} could also be used to derive information related to the star--planet connection as in the case of lithium depletion. The proportions of carbon, oxygen, and rock-forming elements like magnesium, silicon, or iron determines the structure and composition of planets. The C/O elemental number ratio controls the amount of carbides and silicates formed in planets \citep{Larimer75}. 

In the theoretical work by \citet{Bond10} the authors concluded that C/O and Mg/Si elemental ratios are important in  determining the mineralogy of extrasolar terrestrial planets. The work shows that if the C/O ratio is greater than 0.8 (under the assumption of equilibrium), Si exists in a solid form primarily as carbide, and is a factor of two higher than the solar ratio, C/O$_{\sun}\simeq$ 0.55 \citep{Caffau10}. On the other hand, if this ratio of C/O is below the 0.8 value, Si will form silicates (SiO etc.) and be present in rock-forming minerals. This has led to suggestions that there should exist exoplanets consisting of carbides and graphite instead of Earth-like silicates \citep{Bond10}.

However, as a result of the chemical properties of C the gaseous C/O number ratio in planets can vary from the stellar value depending on different parameters (temperature, pressure, etc.) and processes during planet formation, including the initial location of formation of the planetary embryos, the migration path of the planet, and the evolution of the gas phase of a protoplanetary disk \citep{Thiabaud10a, Thiabaud10b, Madhusudhan16}.

The recent works by \citet{Suarez18} and \citet{Bedell18} aimed to determine the mineralogy of planetary companions looking at C/O number ratios of planet hosts. 
In the work by \citet{Suarez18}, the determined C/O ratios revealed different kinds of planetary systems that can be formed, most of them unlike to the solar system. They found that 100\,\% of their sample of stars with detected planets has C/O$<0.8$. Meanwhile, in a spectroscopic study of Sun-like stars, \citet{Bedell18} found that the ratios of C/O in solar metallicity stars are homogeneous to within 10\,\% throughout the solar neighborhood, implying that exoplanets may exhibit much less diversity in their composition than previously thought. 
The recent work by \citet{Pavlenko19} determined carbon and oxygen abundances in atmospheres of the supersolar metallicity stars with and without detected planets and confirmed that metal-rich dwarf stars with planets are more carbon-rich in comparison with nonplanet host stars, with a statistical significance of 96\,\%. 


In this work, we pay special attention to determine the lithium, carbon, and oxygen abundances in a sample of dwarf stars, mainly of spectral classes F, G, and K, in the solar neighborhood with the aim to provide new key insights in the era of exoplanet surveys. Here we address some of the problems mentioned above by homogeneously determining precise abundances of Li, C, and O and analyzing links between the lithium abundance and stellar parameters (e.g., age, \Teff, metallicity, and rotation) in dwarf stars.
The choice of using only dwarf stars is due to the fact that their atmospheres should present the original chemical composition of their birth places and carbon abundances in the atmospheres of dwarf stars are not affected by internal mixing of material comparing to giants.

 \begin{figure*}
\resizebox{\hsize}{!}{
\includegraphics[width=0.8\columnwidth]{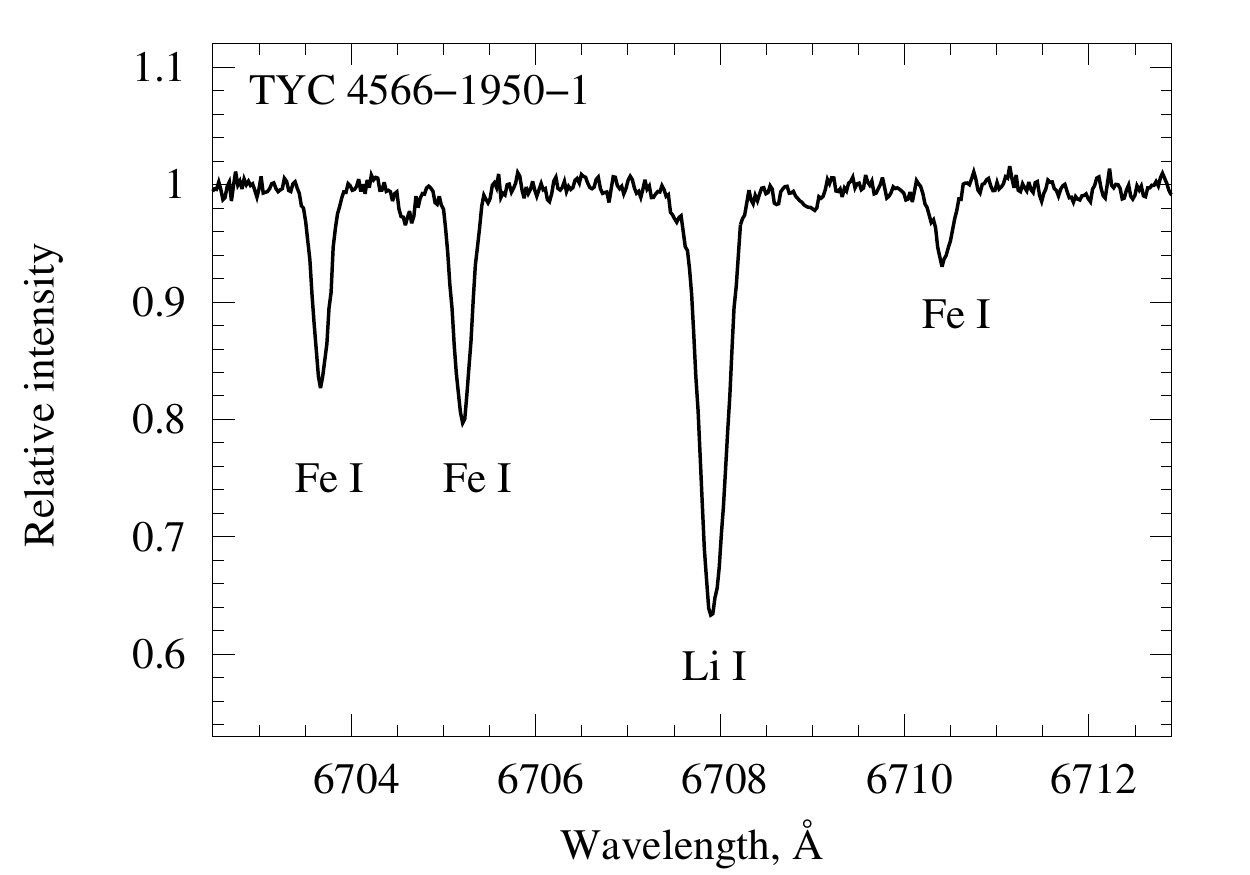}
\hspace{-1.02cm} 
\includegraphics[width=0.8\columnwidth]{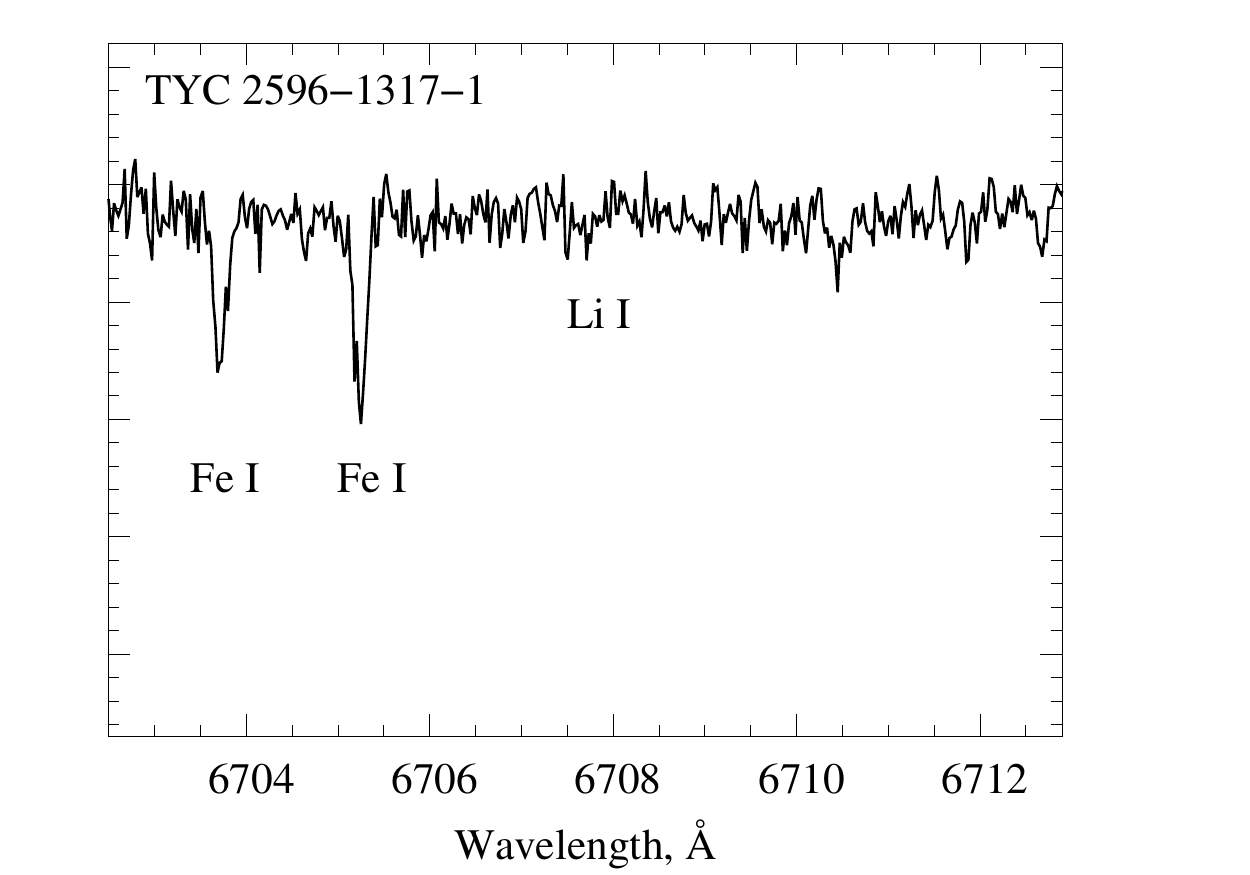}
}
\caption{Examples of observed spectra for stars TYC~4566-1950-1 with the notable lithium line at 6708 \AA~ and TYC~2596-1317-1 with the weak lithium line. 
In the latter and other similar cases only \ALi upper limits had been determined.}
\label{fig:measuplim}
\end{figure*}

\begin{figure*}
	\graphicspath{ {} }
	\includegraphics[width=2\columnwidth]{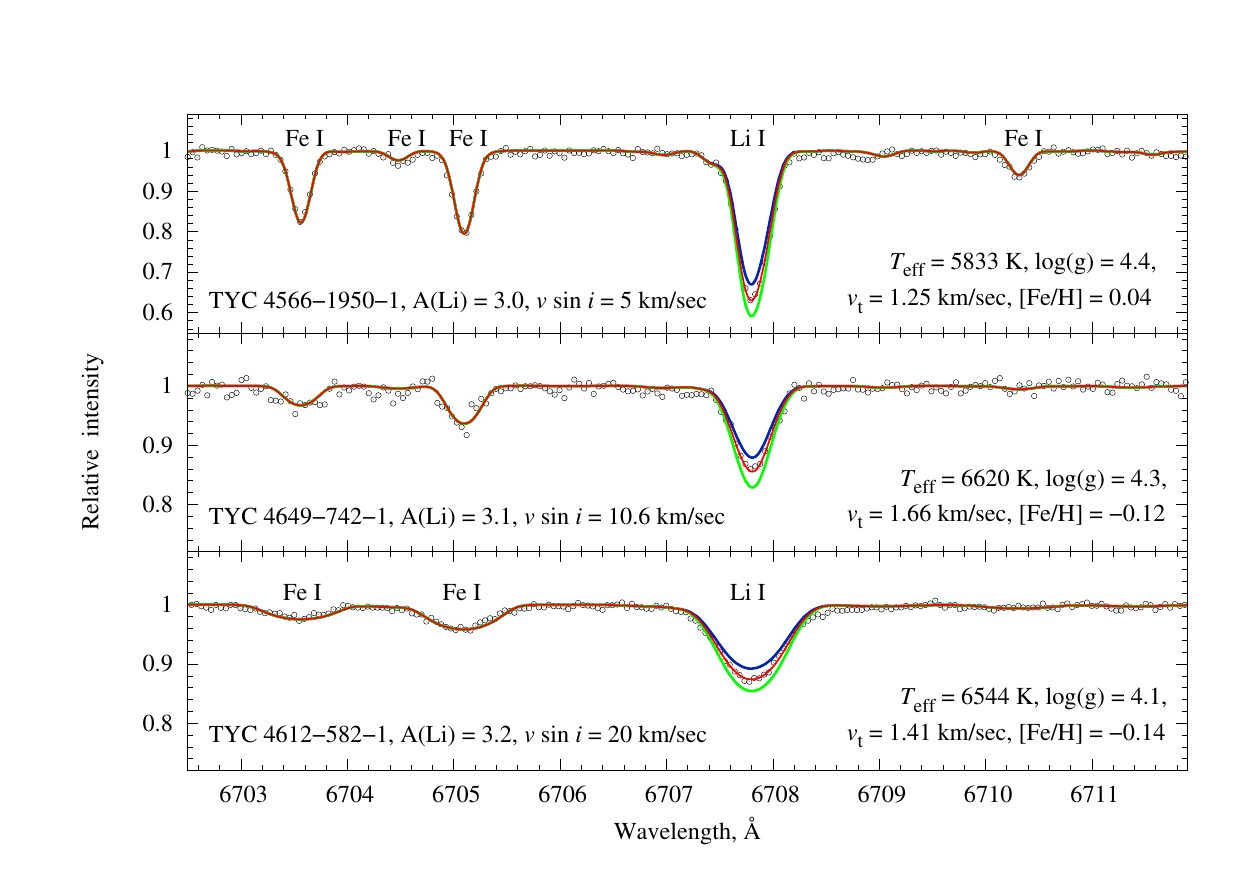}
    \caption{Examples of the observed and synthetic spectra around the lithium 6708 \AA~ line with the lithium abundance \ALi~$\sim$~3~dex. The red solid line shows the best fit while the green and blue solid lines indicate $\pm$0.1 dex. The atmospheric parameters and values of the rotational velocities {\it{v}}\,sin\,{\it{i}} (\kms) are also indicated. }
    \label{fig:lithabmax}
\end{figure*}

\textbf{\begin{figure*}
\resizebox{0.9\hsize}{!}{
\includegraphics[width=0.85\columnwidth]{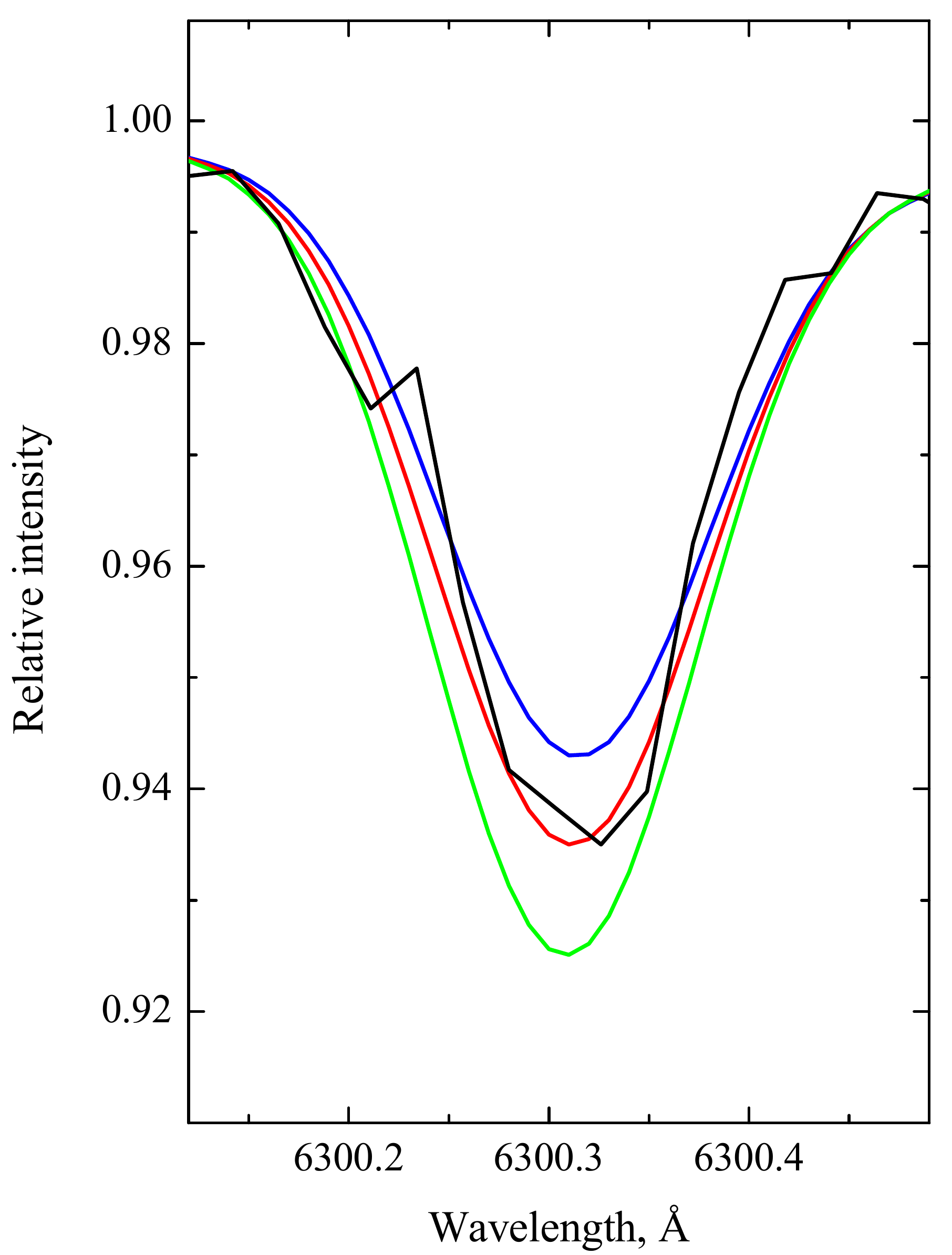}
\includegraphics[width=0.83\columnwidth]{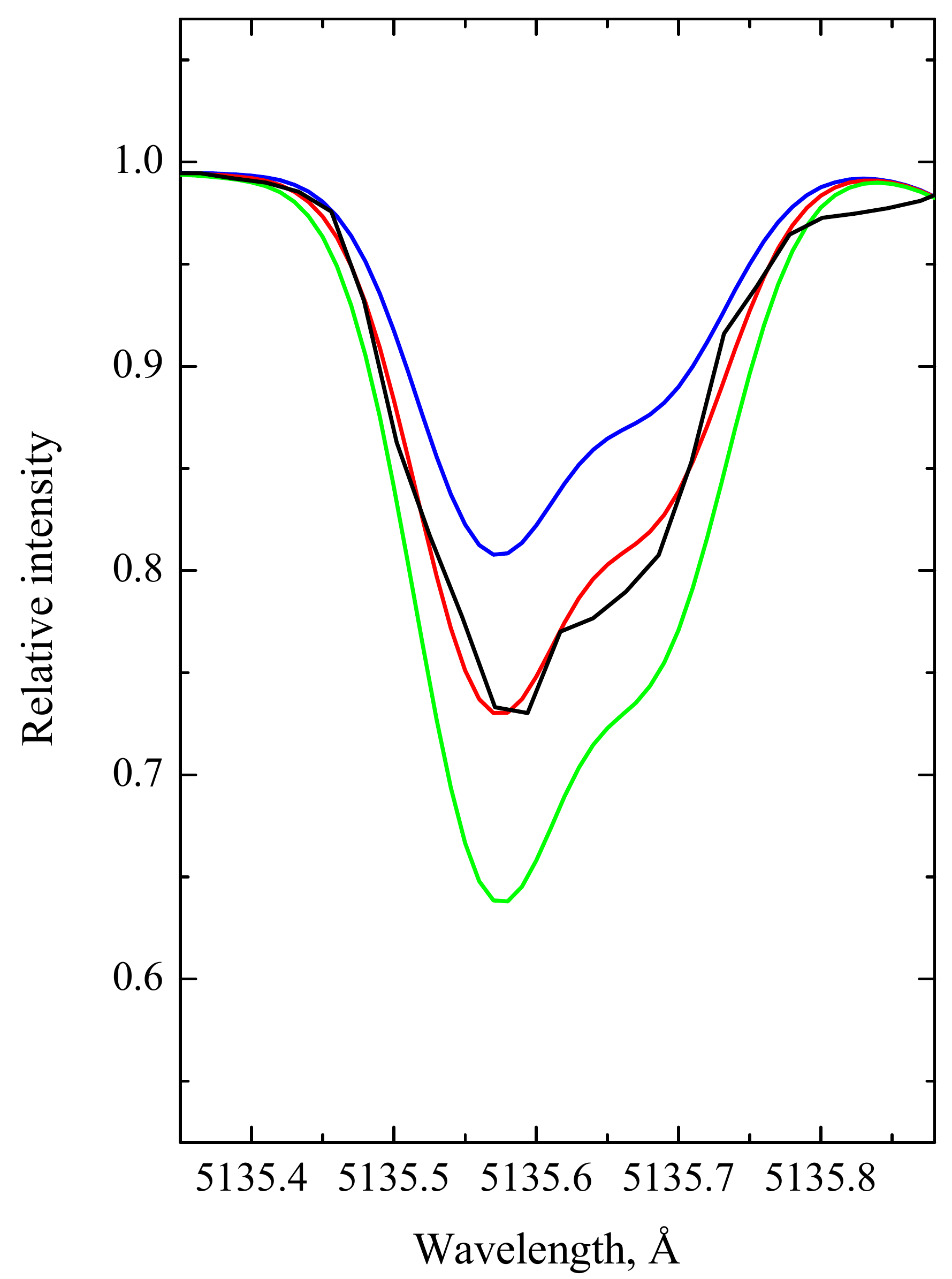}
\includegraphics[width=0.818\columnwidth]{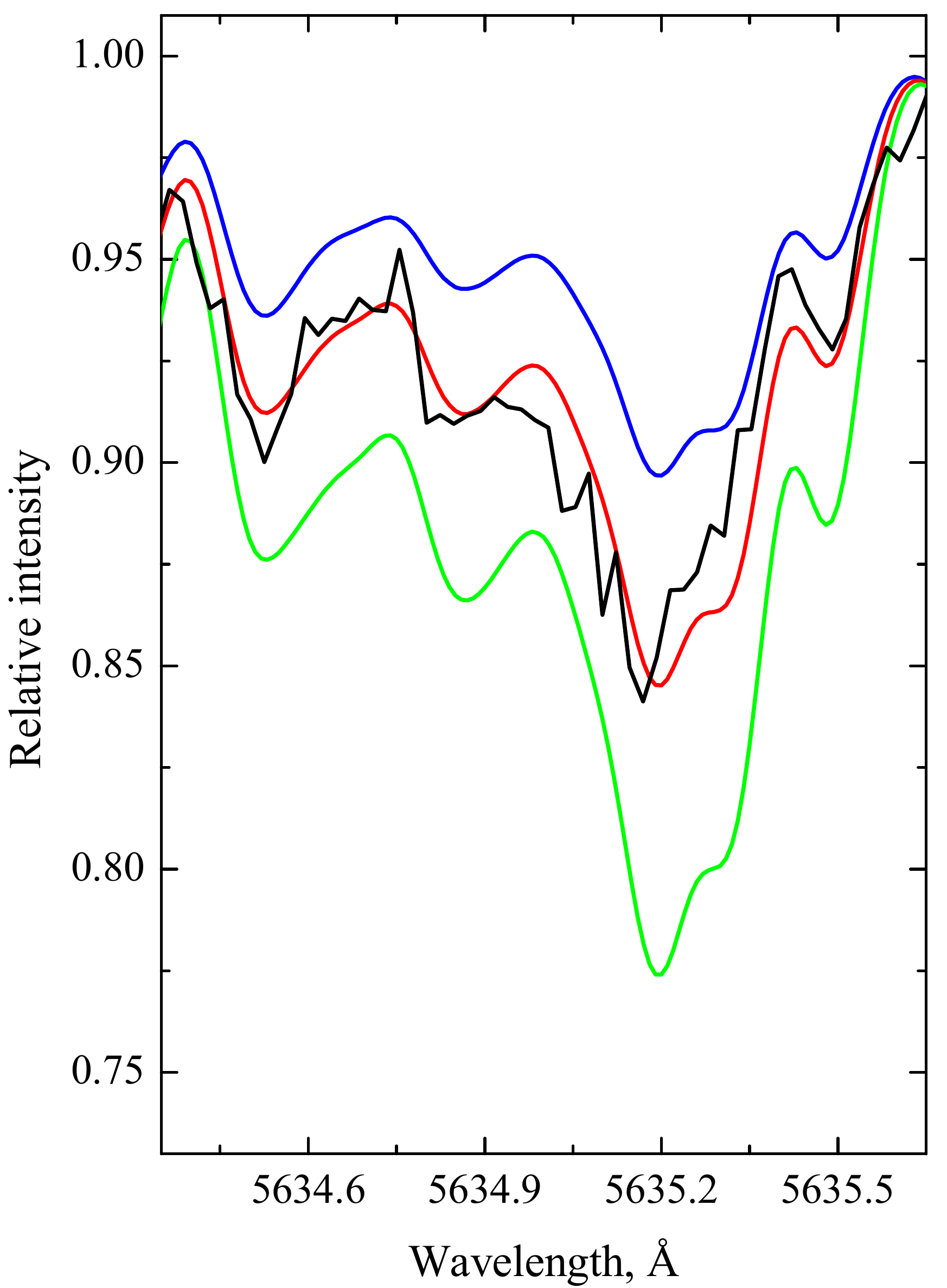}
}
\caption{Examples of the synthetic spectrum fits to the forbidden [O\,{\sc i}] line at 6300~{\AA} and to the ${\rm C}_2$ Swan (1, 0) band head at 5135~{\AA}, and to the ${\rm C}_2$ Swan (0, 1) band head at 5635~{\AA} in the star TYC~4493-620-1. The observed spectra of the star TYC~4493-620-1 are shown as the black solid lines. The red solid lines represent the best fit with [C/Fe]=0.05\,dex and [O/Fe]=0.02\,dex, while the blue and green solid lines represent a change in abundance by $\pm0.1$ dex to the corresponding elements.}
\label{fig:synthetic}
\end{figure*}}

\section{Stellar sample and analysis}\label{sec:Sample and analysis}

\subsection{Observational Data Set}\label{sec:Observational data set}
This work is based on spectroscopic data from the Spectroscopic and Photometric Survey of the Northern Sky (SPFOT) project \cite{Mikolaitis18, Mikolaitis19}. The SPFOT survey aims to provide the detailed chemical composition and the main spectroscopic parameters (e.g., \Teff, \logg, and \FeH)  for bright stars ($V<8$~mag) in the northern sky.  Spectral observations for this project were  carried out with the 1.65\,m telescope at the Mol\.{e}tai astronomical observatory of Vilnius University in Lithuania and the high-resolution Vilnius University Echelle Spectrograph \citep{Jurgenson2016} covering a wavelength range from 4000 to 9000~\AA. From this survey, we took a data set consisting of 249 F, G, and K spectral type dwarf and subgiant stars observed in 2017--2018.  The reader is directed to the papers by \citet{Mikolaitis18,Mikolaitis19} for a more detailed description of the observations, data reduction, determination of the main atmospheric parameters, stellar ages, and detailed abundances for 17 elements (Na, Mg, Al, Si, S, K, Ca, Sc, Ti, Vi, Cr, Mn, Fe, Co, Ni, Cu, and Zn). Like in the previous papers, we used a spectral synthesis method and the code TURBOSPECTRUM \citep{Alvarez1998} with the MARCS stellar atmospheric models \citep{Gustafsson2008}. Unlike in previous papers where abundances of chemical  elements were determined using an automated software, the lithium, carbon, and oxygen abundances we determined visually inspecting the fits for each star. 
The lithium abundance $A$(Li) was determined from profiles of the Li\,I~6708~\AA\ line (if this line was too weak or too noisy, only the upper limit \ALi~was determined), see Figure~\ref{fig:measuplim}. The fits for some Li-rich stars are provided in Figure~\ref{fig:lithabmax}
where the best fit was determined by eye and represented by the (red) middle solid line whereas the other two solid lines indicate $\pm0.1$~dex.

The nonlocal thermodynamic equilibrium (NLTE) effects are negligible for the majority of stars in our sample and have not been taken into account, as for  stars with larger lithium abundances they would be lower than 0.1~dex, thus the results are valid within the assumptions of 1D model atmospheres in LTE and hydrostatic equilibrium \citep[see, e.g.][]{Lind09}.  

For the carbon abundance determination we used two regions: ${\rm C}_2$ Swan (1, 0) band head at 5135~{\AA} and ${\rm C}_2$ Swan (0, 1) band head at 5635~{\AA}. The oxygen abundance was determined from the forbidden [O\,{\sc i}] line at 6300~\AA\ (Figure~\ref{fig:synthetic}). The oscillator strength values for \textsuperscript{58}Ni and \textsuperscript{60}Ni, which blend the oxygen line, were taken from \citet{Johansson03}. The NLTE effects for C$_2$ Swan bands have not been investigated in detail; however it is argued that NLTE effects should not affect the abundances significantly since the forbidden [C\,{\sc i}] line and ${\rm C}_2$ lines give the same carbon abundances \citep{Gustafsson99}. On the other hand, the forbidden oxygen line at 6300.3~\AA\ has been studied extensively, and it is concluded that this line can be perfectly described in LTE \citep{Asplund2005}. 

All  the  synthetic  spectra  have  been  calibrated  to  the  solar spectrum by \citet{Kurucz05} with log~$A_{\odot}$(Li) = 1.05, log~$A_{\odot}$(C)~=~8.39, and log~$A_{\odot}$(O) = 8.66 taken from \citet{Grevesse07}. Several examples of the synthetic spectra fits for C and O lines are presented in Figure~\ref{fig:synthetic}.

\begin{table*}
 \caption{ The Median Values of Stellar Atmospheric Parameters and Their Errors.}
 \label{tab:median}
 \begin{tabular}{ccccccc}
  \hline
  \hline
  $T_{\rm eff}$ $|$ $\sigma$  & \logg $|$ $\sigma$  & [Fe/H] $|$ $\sigma$  & $v_{\rm t}$ $|$ $\sigma$ & $A$(Li) $|$ $\sigma$  & [C/H] $|$ $\sigma$   &  [O/H] $|$ $\sigma$   \\
 (K) &   &  & (\kms)  &  &  &  \\
  \hline
6064 $|$  $\pm46$ & 4.22 $|$  $\pm0.30$ & $-0.10$ $|$  $\pm0.11$ & 1.04 $|$  $\pm0.27$ & 2.44 $|$  $\pm0.04$ & $-0.03$ $|$  $\pm0.02$ & 0.07 $|$  $\pm0.14$ \\
  \hline
 \end{tabular}
\end{table*}

\begin{table}
\centering
\caption{Effects on the Derived Chemical Abundances Resulting from the Atmospheric Parameter Uncertainties for the Program Stars.}
 \label{tab:effects}
 \begin{tabular}{ccccc}
  \hline
  \hline
 Species & $\Delta{T_{\rm eff}}$  & $\Delta{\rm log}\,g$  & $\Delta$[Fe/H] & $\Delta v_{\rm t}$  \\
 & $\pm46 $ K & $\pm0.30$ dex & $\pm0.11$ dex  & $\pm0.27 $ \kms \\
  \hline
$A$(Li) & $\pm0.05$ &  $\pm0.00$  & $\mp0.00$ & $\pm0.01$ \\
C (C$_{2}$) & $\mp0.02$ &  $\pm0.01$  & $\mp0.00$ & $\pm0.00 $ \\
O {[O\,{\sc i}]  } & $\pm0.02$ &  $\pm0.14$  & $\pm0.01$ & $\pm0.01$ \\
 \hline
 \end{tabular}
\end{table}

\subsection{Uncertainties}\label{sec:Uncertainties}
There are two sources of errors in abundances determined in this work: 1) uncertainties caused by analysis of individual lines, including random errors of atomic data or continuum placement and signal-to-noise (S/N) ratio, that affect a single line, and 2) uncertainties in the stellar parameters, that affect all lines together. Uncertainties coming from the atomic data are mostly eliminated because of the differential analysis relative to the Sun. The calculated medians of atmospheric parameter determination errors from all the stars in our sample are presented in Table~\ref{tab:median}. In this table medians of Li, C, and O abundance determination errors are introduced as well.

\begin{figure*}
\resizebox{\hsize}{!}{
\includegraphics[width=0.7\columnwidth]{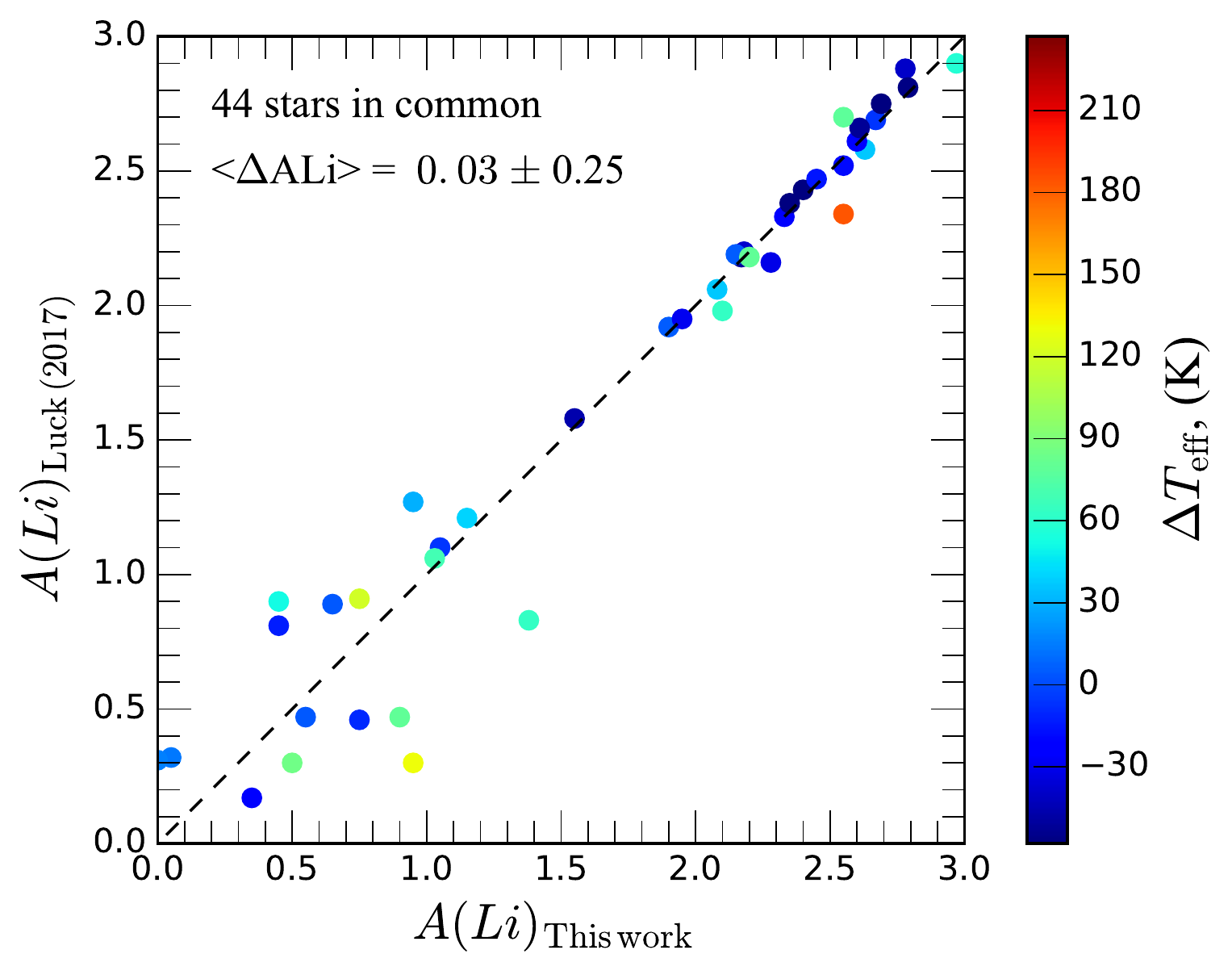} 
\includegraphics[width=0.7\columnwidth]{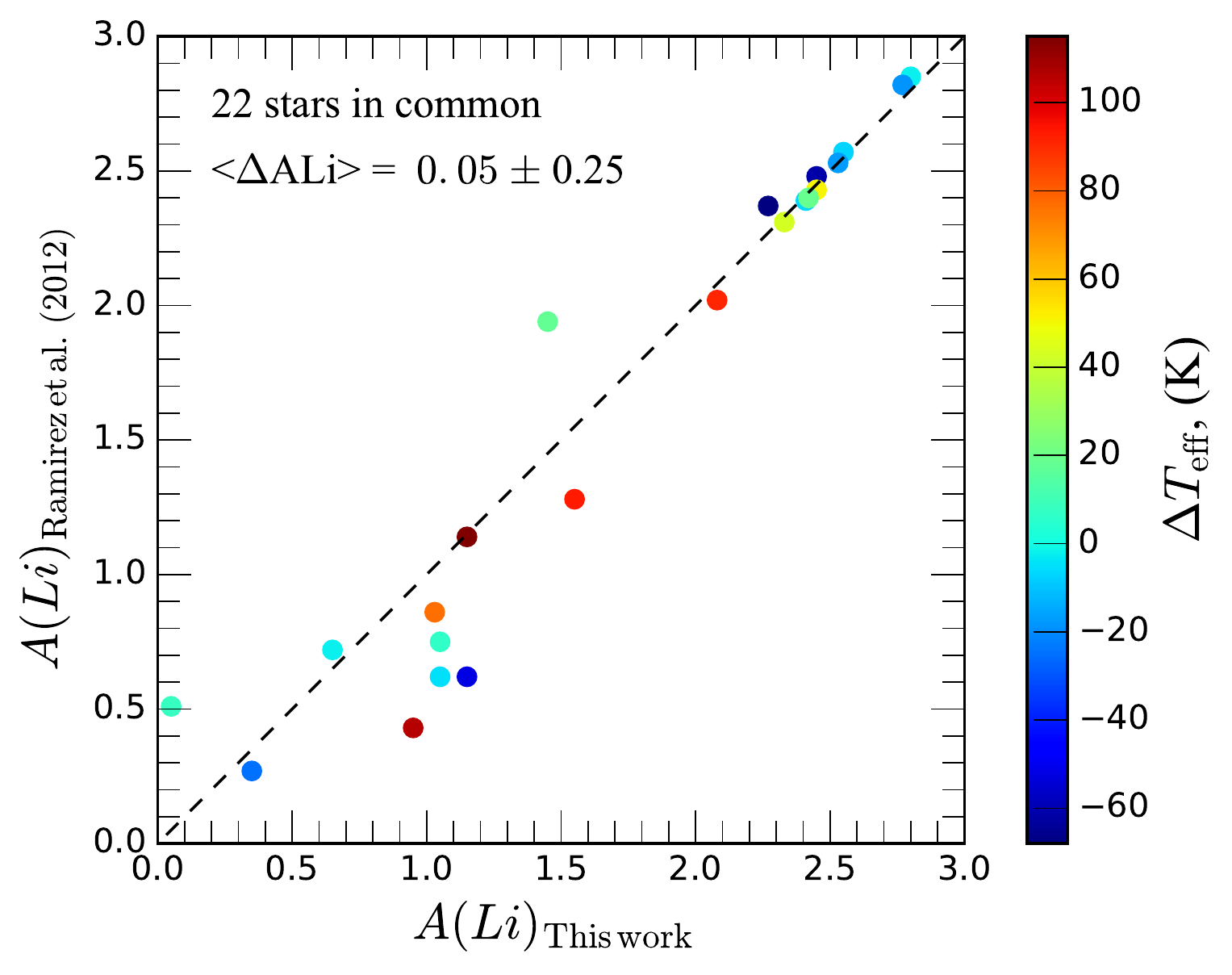}
}
\caption{Comparison plots of \ALi~abundances from this work against those of \citet{Luck17} and \citet{Ramirez12}. The dashed-line shows a 1:1 correlation. The average differences and standard deviations ($\Delta$~\ALi~and~$\Delta$~\Teff) are calculated as our values minus those of other authors. The stars have been color-coded according to a difference in the effective temperatures.}
\label{fig:Li_compa}
\end{figure*}

\begin{figure*}
\resizebox{\hsize}{!}{
\includegraphics[width=0.7\columnwidth]{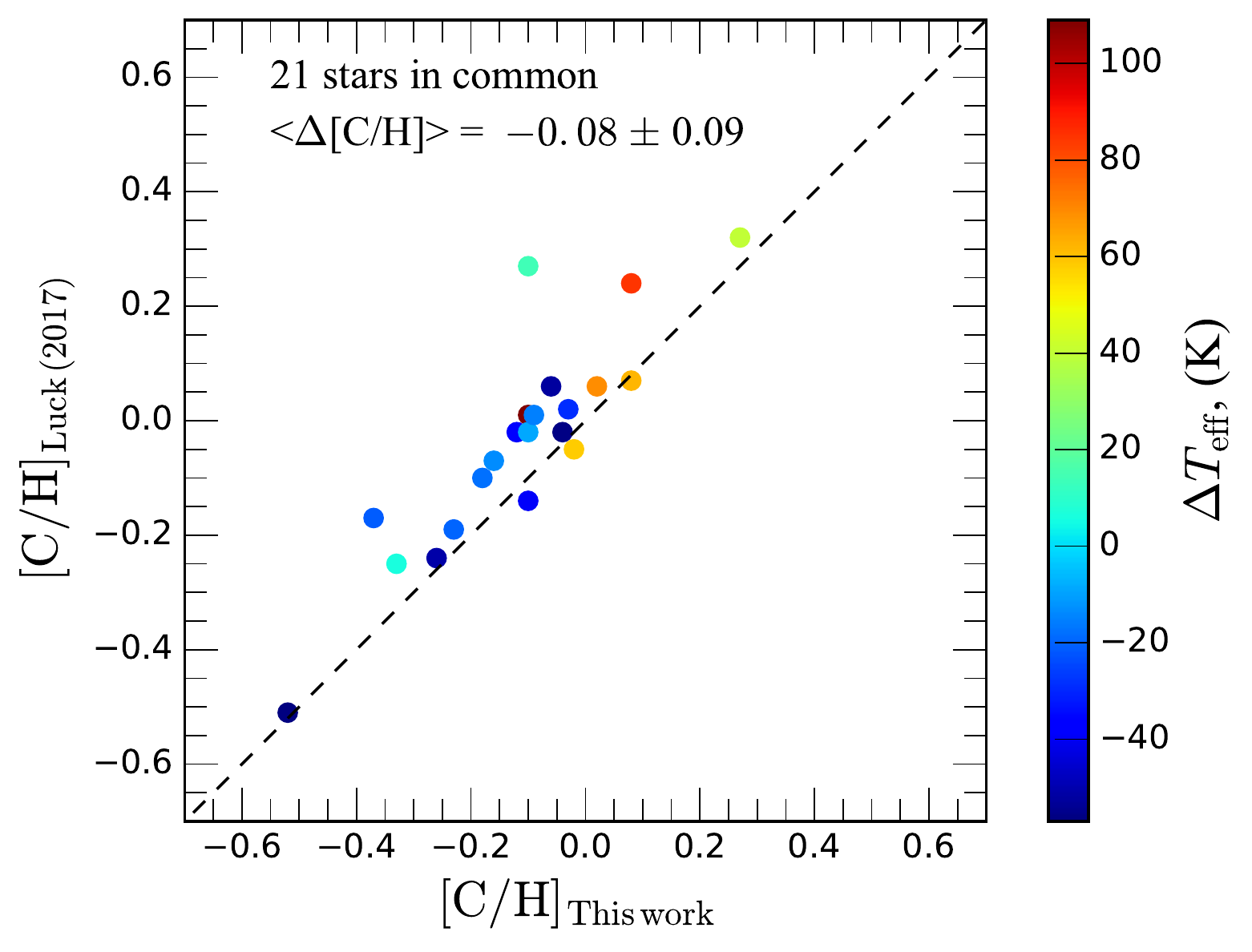}
\includegraphics[width=0.7\columnwidth]{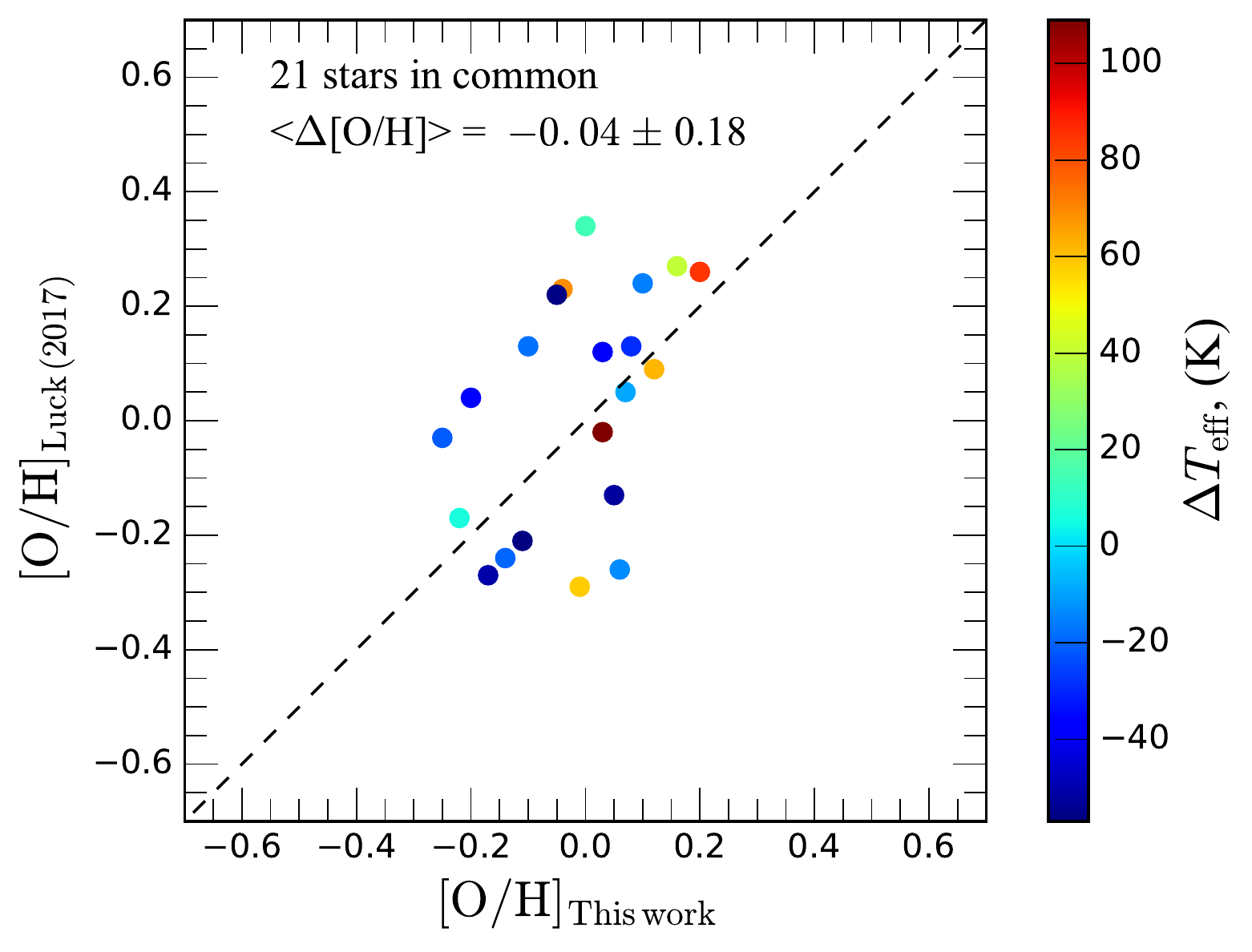}
}
\caption{Comparison plots of [C/H] and [O/H] from this work against the work by \citet{Luck17}. The dashed-line shows a 1:1 correlation. The average differences and standard deviations ($\Delta$~[Element/H]~and~$\Delta$~\Teff) are calculated as our values minus the comparison values. The stars have been color-coded according to a difference in the effective temperatures. }
\label{fig:CO_compa}
\end{figure*}

We calculated changes in abundances caused by the error of each individual atmospheric parameter, keeping other parameters fixed. Results are presented in Table~\ref{tab:effects}. As can be seen, the abundances are not very  sensitive to the changes of atmospheric parameters, only the oxygen abundance sensitivity to the surface gravity is more visible. 

Carbon and oxygen are bound together by the molecular equilibrium, therefore we measured abundances of these two elements in unison, even though carbon and oxygen abundances in dwarf stars are not very sensitive to each other.  Exempli gratia in TYC\,4349-205-1, $\Delta$[C/H] = $\pm 0.10$ causes $\Delta$[O/H] = 0.00 and $\Delta$C/O = $\pm 0.10$; $\Delta$[O/H] = $\pm 0.10$ causes $\Delta$[C/H] = $\pm 0.03$ and $\Delta$C/O = $\pm 0.06$. Thus changes in one element influence abundances in the other very little.

\begin{table*}
	\centering
    \caption{Abundances of Li, C, and O, Atmospheric and Other Parameters of the Sample Stars.}
 \begin{tabular}{lccccccccccccc}
  \hline
  \hline
  Star$^{a}$ & \ALi & {\rm \textit{M}$|$\textit{U}} &\Vsini & [C/H] & \textit{e}\_[C/H] & n$_{\rm C}$ & [O/H] & \textit{e}\_[O/H] & n$_{\rm O}$  & C/O  & Thin$|$Thick & Planet&...$^{a}$\\
 & (dex) & & (\kms) & (dex) & (dex) & & (dex) & (dex) && & & &\\
  \hline
  
      \noalign{\smallskip}
	TYC1547-367-1	&$	0.86	$&	\textit{U}	&$	6.0	$&	& &	&		&		&		&		&	0	&	0	 &...\\
	TYC1563-3551-1	&$	0.77	$&	\textit{U}	&$	1.0	$&	& &	&		&	&		&		&$	1	$&$	0	$ &...\\
	TYC1563-3552-1	&$	2.87	$&	\textit{U}	&$	11.0	$&	& &	&		&		&		&		&$	0	$&$	0	$ &...\\
	TYC2057-709-1	&$	1.98	$&	\textit{M}	&$	1.2	$&$	-0.08	$&$	0.09	$& $2$ & $0.15$ & 0.06 & 1&	0.32	&	0	&	0	 &...\\
	TYC2070-1061-1	&$	2.49	$&	\textit{M}	&$	2.0	$&	& &	&		&		&		&		&$	0	$&	0	 &...\\
    ...&                ...&    ...&    ...&    ...& ...& ...&    ...&    ...&    ...&   ...&    ...&    ...&...\\

\hline
\end{tabular}
\flushleft

$^{a}${\scriptsize
For column names and details see the~Appendix.}\\
(This table is available in its entirety in machine-readable form.)
\label{tab:results}
\end{table*}

\subsection{Comparison with Other Studies}\label{sec:Comparison to other studies}

Figures~\ref{fig:Li_compa} and \ref{fig:CO_compa} show comparisons between the Li, C, and O abundances determined in this work and from two other studies: \citet{Luck17} and \citet{Ramirez12} have 44 and 22 stars in common with our sample for \Li, respectively;  and \citet{Luck17} have 21 stars in common with our sample for the carbon and oxygen abundances. The points are color-coded by the difference of the determined effective temperatures for the same stars. The agreements between the studies are good, on average our \ALi~abundances are about 0.04~dex higher than the \citet{Luck17} and \citet{Ramirez12} values, with a one-sigma dispersion of 0.25~dex. The higher Li values agree better, as the stronger lines provide a more accurate determination, whereas smaller Li values are usually determined from lower S/N spectra and weaker lines which result in only an upper limit determination or larger uncertainties in the values. Those studies use a different technique for the parameter and abundance determinations from the method used in our work. The temperatures in the works by \citet{Luck17} and \citet{Ramirez12} were derived using the \citet{Casagrande10} effective temperature calibration. It is encouraging that the effective temperatures in our and other works are in good agreement: comparing with \cite{Luck17} we obtain on average 20~K higher temperatures with a mean standard deviation of 70~K, and comparing with \cite{Ramirez12} we get on average 16~K higher temperatures with a 50~K mean standard deviation. 

Our [C/H] and [O/H] abundance values are on average about 0.08 and 0.04~dex lower with dispersions of 0.09 and 0.18~dex,  respectively, than the \citet{Luck17} values for a sample of 21 common stars. 
We would like to note, that for the comparison with Luck's~(\citeyear{Luck17}) work we use [O\,{\sc i}] and C$_2$  data per line found in their Table~5. Their Table~5 contains absolute values for carbon abundance from the C$_2$ Swan line primary indicator at 5135~\AA~and oxygen abundances from the forbidden line at 6300~\AA, the same lines that we used for the carbon and oxygen abundance determinations. We decided to scale their absolute values for C$_2$, oxygen and iron (Fe\,{\sc i}) using the solar values from \citet{Grevesse07} as in our work to be on the same elemental abundance scale. 

There are several other studies on carbon and oxygen abundance determinations with which we have fewer stars in common. For example, in the work by \citet{Petigura11} we find eight stars in common and our [C/H] abundance values are on average $0.05\pm 0.10$~dex lower than theirs. However, we see no difference for [O/H] abundance determination between the two studies.

\section{Results and Discussion}\label{sec:Results}

Table~\ref{tab:results} shows an example of the online table containing the resulting lithium, carbon, and oxygen abundances, C/O number ratios as well as atmospheric parameters, whether the star belongs to the thin or thick disk, and other information on the stars in our analyzed sample. For column names and details see the Appendix.

Our stars have metallicities, effective temperatures, and ages between \mbox{(--0.7 $\div$ 0.4) dex}, \mbox{(5000 $\div$ 6900) K}, \mbox{(1 $\div$ 12) Gyr}, accordingly. 
Accurately measurable lithium lines were found in spectra of 149 stars, the upper limits were determined for the other 100 stars (60\,\% and 40\,\% of the sample, respectively). The largest fraction of stars has \ALi~$\geq 2.0$~dex and makes up $\approx$ 51\,\% of the sample. Carbon and oxygen abundances were determined for 78 stars. 

\begin{figure}
\center{\includegraphics[width=1.0\columnwidth,angle=0]{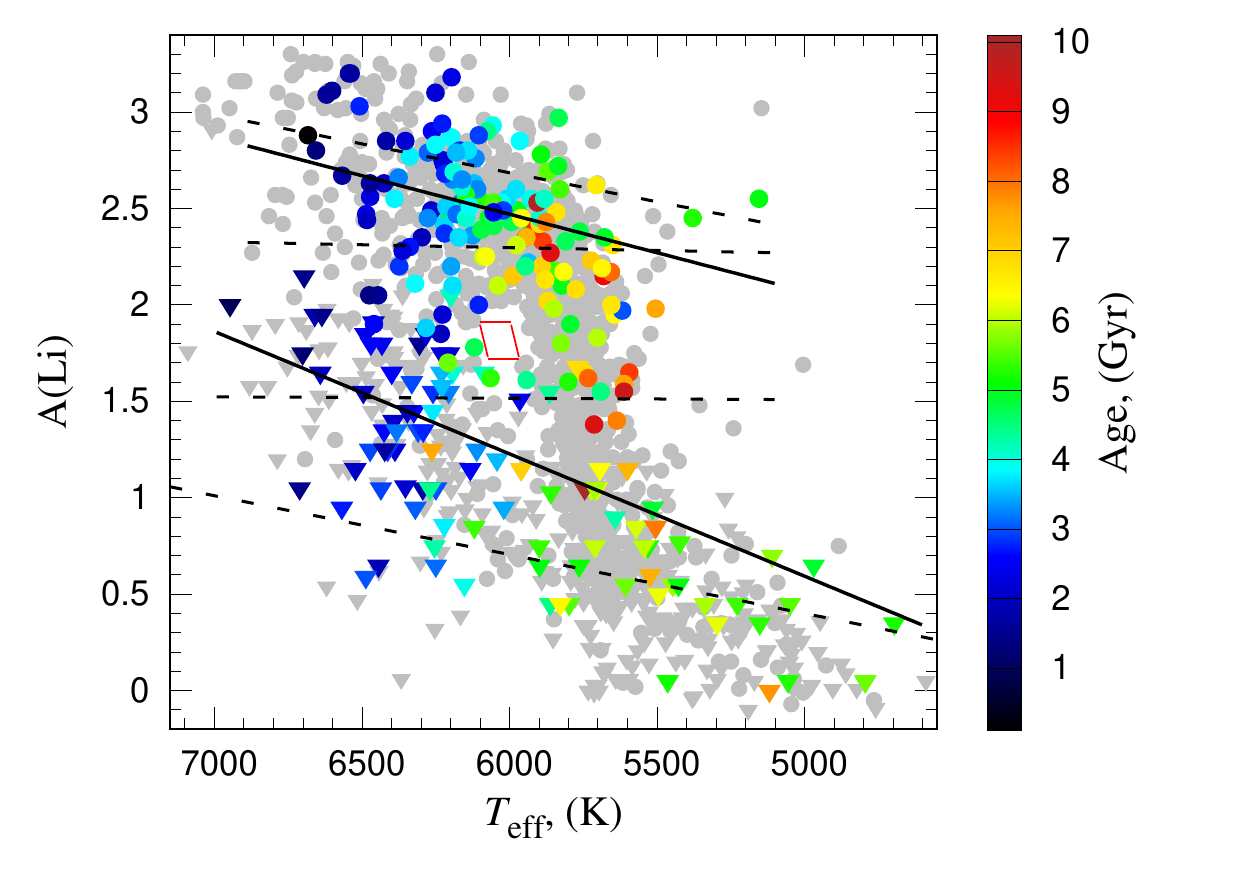} }
\caption{Abundance of lithium \ALi~as a function of effective temperature. 
The stars with measurable \ALi~and upper limits are shown as circles or triangles, respectively, and color-coded by their ages.
For the comparison, we plotted results from \citet{Ramirez12} as gray circles and triangles.
The linear approximations of \ALi~as a function of \Teff~are shown as the black solid lines for two groups of stars with \ALi~$\geq 2$ and \ALi$< 2$.
The dashed lines represent similar fits for the ranges \ALi$\geq 2.5$; $2.0 \leq$ \ALi$<2.5$; $1.0 \leq$~\ALi$<2.0$, and \ALi$<1.0$.
The quadrangle with the red perimeter shows an approximate location of the lithium desert near \Teff $\approx$ 6025~K and \ALi~$\approx 1.8$.
}
\label{fig:ALiTAge}
\end{figure}

\begin{figure}

\begin{minipage}[]{\linewidth}
\center{\includegraphics[width=1.0\linewidth]{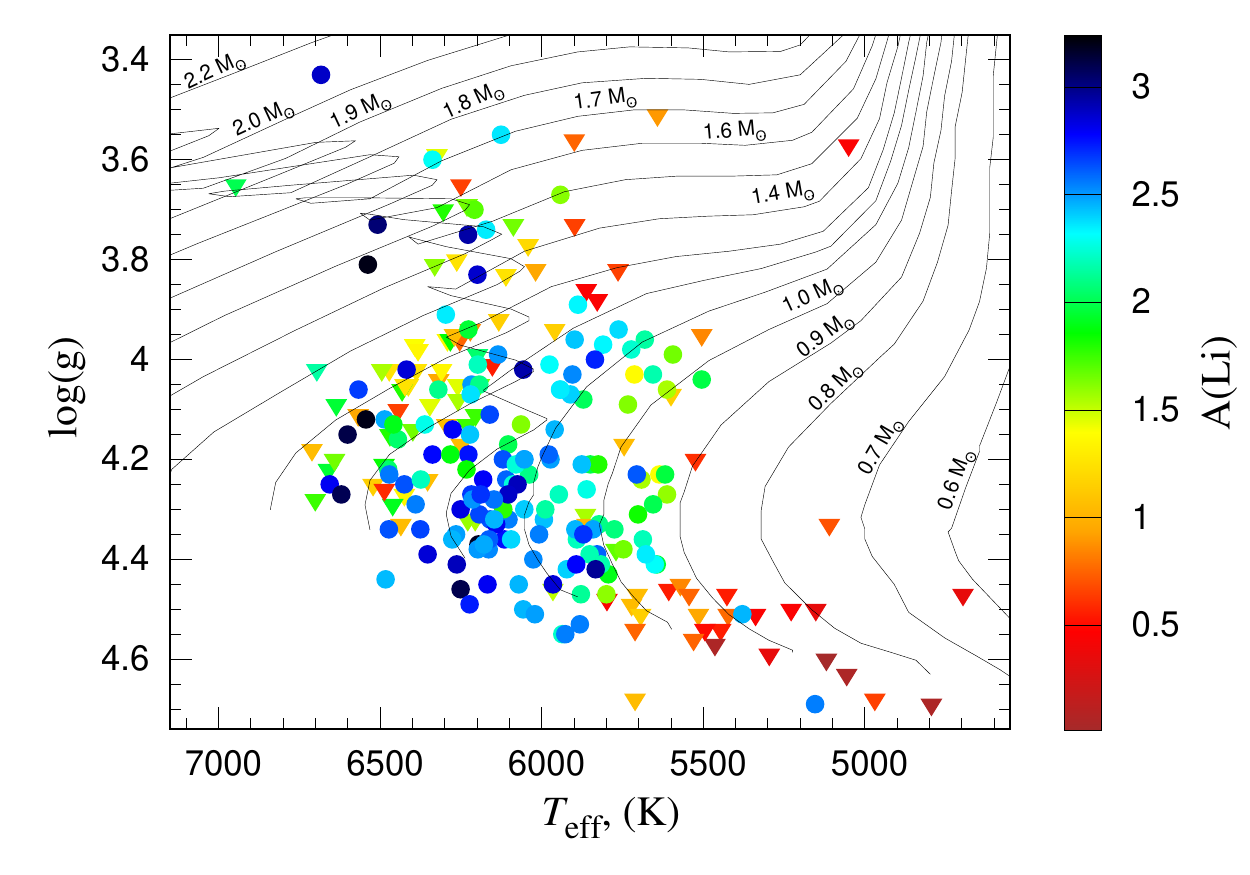}}
\end{minipage}
\caption{HR diagram in a surface gravity versus effective temperature plane. Stars are color-coded according to their \ALi.
The black lines show the evolutionary tracks, which were taken from the work by \cite{Girardi00} with masses $(0.6\div 2.2)$\MS and the fixed $Z_{\rm ini} = 0.019$. }
\label{fig:loggTALi}
\end{figure}

\subsection{Lithium Abundances versus Effective Temperature}
\label{sec:LivsTeffLoggFeH}

In Figure~\ref{fig:ALiTAge} we show the distribution of lithium abundance \ALi~as a function of effective temperature \Teff. Stars from our sample are color-coded by their ages. 
The lithium abundance versus effective temperature plane could be divided into six regions of interest:
1) the region of lithium-rich stars with \ALi $\geq$ 2.0; 2) the area of Li-depletion at \mbox{5500~K $\leq$ \Teff $\leq 5950$~K}; 3) the region of the lithium-poor stars at \Teff$ > 6000$~K; 4) the area of the lowest values \ALi~$\leq1.0$ at \Teff$<5500$~K; 5) the region of Li-abundant chromospherically active stars with 5200~K$ < $\Teff$ < 5700$~K; and 6) the lithium desert with \ALi~$\approx$~1.8 between 5950~K~$\div$~6100~K.  
For the comparison in Figure~\ref{fig:ALiTAge} we plotted results from the work by \citet{Ramirez12} as gray circles and triangles. This \ALi~versus~\Teff~plane has similar trends found in previous studies \citep[see, e.g.,][]{Ramirez12,LopezValdivia15,Luck17,Bensby18}:

1) Lithium-rich dwarfs area. Stars with \ALi $\geq$ 2.0 are located in a broad ($\Delta$\ALi $\simeq$ 1.0) and a rather vague band that stops at \Teff~$\simeq$~5600~K. The highest values of lithium (\ALi~$\geq$~3) were found for seven stars which are among the hottest and youngest objects in our sample with effective temperatures \Teff~$\geq$~6200~K and ages $\leq$2.8~Gyrs. A linear regression fit to the area of \ALi~$\geq$~2.0 values shows that the slope per 1000~K equals to 0.4~$\pm$~0.1~dex as indicated with the upper solid line in Figure~\ref{fig:ALiTAge}. We find that this value is close to the value of \mbox{$0.45 \pm 0.1$}~dex/1000~K found in the work of local metal-rich stars by \cite{Pavlenko18}.

Next, we divided the stars with
\ALi~$\geq 2.0$ into two bins with abundances 2.0~$\leq$~\ALi$<2.5$ and \ALi~$\geq 2.5$ and determined gradients for both. We obtained that the slope per 1000~K for the 2.0~$\leq$~\ALi$<2.5$ bin is $0.01 \pm 0.07$~dex and for the \ALi~$\geq 2.5$ we got $0.28 \pm 0.09$~dex as indicated by upper dashed lines in Figure~\ref{fig:ALiTAge}. Hence, the correlation between \ALi and \Teff is stronger for the bin where \ALi~$\geq 2.5$ and is similar to the larger bin of \ALi~$\geq 2.0$ while it is almost flat or non-existent for the 2.0~$\leq$~\ALi$<2.5$~bin. 

2) Lithium-depletion area. We note a region of strong lithium depletion towards lower \Teff in the relatively narrow range \mbox{5500~K $\leq$ \Teff $\leq$ 5950~K} (or $M\sim$~($0.9\div1.1$)~\MS). As it is known, this depletion is due to a significant increase of convective depths of such stars since the deeper convection transports lithium to temperatures where it is destroyed. The mean age of these stars in our sample is $6.3 \pm 1.5$~Gyr.

3) Lithium-poor stars with \Teff$>6000$~K. Opposite to the mentioned area 2) we observe a fraction of dwarf stars with \ALi$<2$ with the effective temperature \Teff~$>$~6000~K, however, the majority of dwarfs have only \ALi upper limits and the mean age of these stars is equal to $2.9 \pm 1.15$~Gyr. 

4) Stars with \ALi~$\leq$~1.0 at \Teff$<5500$~K.
Moving towards the lower \Teff$<5500$~K, we see a number of lithium-poor stars with \ALi~$\leq$~1. Some of stars have only upper limit values decreasing to \ALi~$\approx$~0. 
In this region (\ALi~$\leq$~1.0 at \Teff$<5500$~K) lithium abundances reach a plateau which exist because the convection depth in such stars is not growing any more with decreasing temperatures (or masses) \citep[see e.g.,][]{Luck17}. 

5) Li-abundant chromospherically active stars with 5200~K~$<$\Teff$<5700$~K. This region was identified by \citet{Mishenina12} while investigating a correlation between \ALi and chromospheric activity of stars. We found two chromospherically active stars with $M\sim$1\MS, \FeH$\sim-0.1$ and \ALi~$\sim$~2.5 in this temperature region, while other stars with similar atmospheric parameters had \ALi~$\sim$~0.5. Chromospheric activity of  
TYC-4566-120-1 ($\pi^1$~UMi~B, HD~139813) was investigated by e.g., \citet{Isaacson10}, \citet[and references therein]{Eisenbeiss13},
and TYC-4609-535-1 (V368~Cep, HD~220140) was investigated by e.g., \citet[and references therein]{Zhang15}. Thus, we confirm the conclusion that chromospherically active stars may have quite large lithium abundances in this effective temperature interval. 

6) Lithium-desert area. Lastly, the astonishing region of the so-called lithium-desert is clearly visible in the effective temperature range of 5950~K~$\div$~6100~K with \ALi~$\approx$~1.8 (see the red quadrangle in Figure~\ref{fig:ALiTAge}). According to \citet{Aguilera-Gomez18}, stars below the lithium-desert have evolved from the lithium dip. A dip-like feature appears in \ALi~versus~\Teff~plane around 6700~K, first seen in old clusters and less evident in field stars. These stars are more massive, more evolved and have lower \ALi~abundances compared to other stars of the same temperature, which are located above the desert and have higher Li abundances.

Looking at the Li abundance correlation with effective temperature for stars with \ALi$<2$, which is indicated in Figure~\ref{fig:ALiTAge} with a lower solid line, we see that the overall slope is quite similar to the one for stars with larger Li abundances (\ALi~$\geq$~2). 
The \ALi slope per 1000~K equals to $0.6 \pm 0.1$. A corresponding slope found by \cite{Pavlenko18}  is a bit larger and equals to $0.85 \pm 0.18$~dex but has a larger error. We divided the stellar sample with \ALi$<2$ in two bins as well. The slopes for 1~$\leq$~\ALi$<2$ and \ALi$<1$ bins per 1000~K are equal to $0.01 \pm 0.11$~dex and $0.31 \pm 0.07$~dex, respectively. Hence, the correlation between \ALi~and \Teff~is stronger for the \ALi$<1$ bin and almost non-existent for the 1~$\leq$~\ALi$<2$ bin (indicated by two lower dashed lines in Figure~\ref{fig:ALiTAge}).

In Figure~\ref{fig:loggTALi}, we show investigated stars in the temperature--gravity plane. The stars with measurable \ALi~and upper limits are shown as circles and triangles, respectively, and color-coded by their \ALi.
The black-lines indicate evolutionary tracks, which are taken from the work by \cite{Girardi00} with masses (0.6 $\div$ 2.2)\MS and fixed $Z_{\rm ini}$ = 0.019. 
We see that the lowest \ALi dwarfs are on the lower right part of the main sequence (MS) in the area of lowest \Teff~of the sample, whereas the highest Li abundances are spread in the region of the MS with highest \Teff. 
\ALi~is predominantly higher among hotter and more massive stars. This main trend is in agreement with results of previous studies \citep[see, e.g.,][]{Ramirez12,Luck17,Bensby18,Fu18,Pavlenko18}. 

\begin{table*}
 \caption{The Stars with Similar Main Parameters but Different Lithium Abundances.}
 \label{tab:remarkably}
 \begin{tabular}{lcccccccc}
  \hline
  \hline
 { \raisebox{-2.0ex}[0cm][0cm]{Star}} & \Teff & \logg & \FeH  & \vt   & \Vsini & Age  & \ALi   \\
      &  (K) &     &  &(\kms) & (\kms) & (Gyr) &  \\
 \hline   
TYC4612-582-1	& 6544 & 4.1 &	 $-0.14$  &   1.41  & 20  	& 1.7    & 3.20 \\
TYC3890-1396-1  & 6445 & 4.2 &    0.13   &   1.59  & 25     & 1.6    & 2.05 \\

TYC4426-975-1	& 6486 & 4.1 &	$-0.20$   &   1.39  & 9      & 2.4    & 2.47 \\
TYC4488-687-1	& 6459 & 4.1 &	$-0.23$   &	1.46   & 8   	& 2.5    & 1.90  \\

TYC3132-2016-1	& 6228 & 3.8  &	0.00	& 	1.60  &	11      & 2.7	 & 2.94  \\	 
TYC4414-1663-1	& 6227 & 3.9  &	$-0.11$	& 	1.83  &	14      & 2.2	 & 1.95  \\	       

TYC2661-155-1	& 6297 & 3.9  &	0.26	& 	1.54  & 9	    & 1.9	 & 2.35  \\
TYC2592-1547-1	& 6294 & 4.1  &	0.28	& 	1.57  & 12      & 1.7	 & 1.05  \\ 

TYC4566-1950-1	& 5833 & 4.4 &	 0.04   &	1.25  & 5.0	    & 4.7    & 2.97   \\
TYC2639-2460-1  & 5817 & 4.4 & $-0.21$   &   0.77  & 3.0     & 6.3    & 2.17   \\ 
TYC2613-2218-1  & 5794 & 4.4 &   0.10   &   0.79  & 0.1     & 4.9    & 1.90  \\

TYC4409-1023-1	& 5706 & 4.2  &	0.07	& 	0.83  & 2	    & 6.6	 & 2.62  \\
TYC4366-1351-1	& 5691 & 4.2  &	0.14	& 	0.83  & 1       & 4.5	 & 1.55  \\ 
 \hline
 \end{tabular}
\end{table*}

However, there are stars with the same atmospheric parameters and ages but different lithium abundances. We show some examples in Table~\ref{tab:remarkably}. Within uncertainties, those stars have similar ages, atmospheric parameters, and rotational velocities, while their values of \ALi differ significantly. It could mean that in addition to the abovementioned main parameters (\Teff, mass, age, etc.) lithium abundances in the stellar atmospheres can be affected by external factors such as presence of planets, planet migration, or inside-out disk formation and stellar radial migration \citep[see][and references therein]{Chen06, Prantzos17,Guiglion19}.

\subsection{Lithium Abundances versus Metallicity}

\begin{figure}
\center{\includegraphics[width=\linewidth]{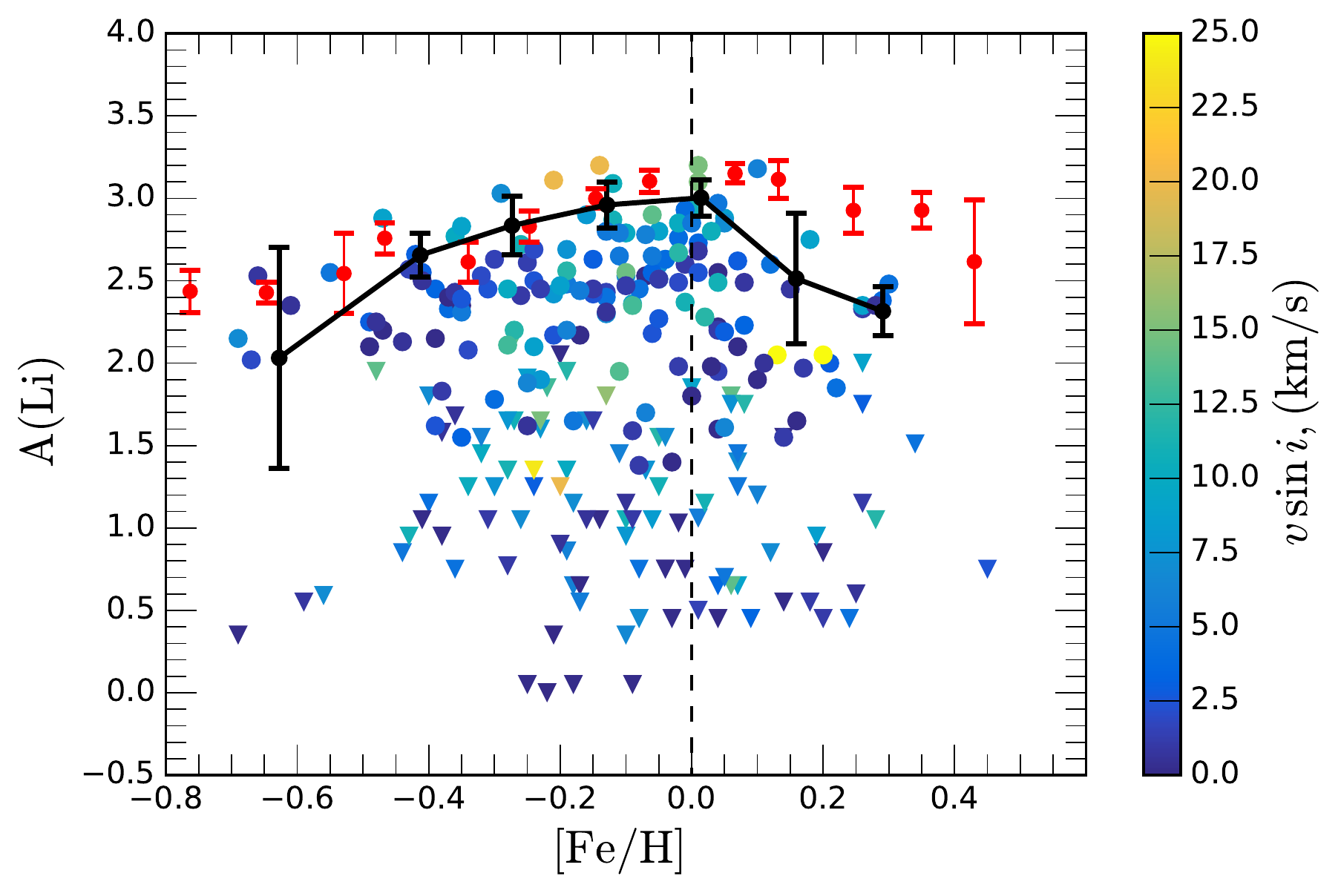} }
\caption{Abundances of lithium, \ALi~as a function of metallicity [Fe/H]. Our results (filled circles -- measurable and triangles -- upper limits) are color-coded according to their rotation \Vsini. The black dots indicate the mean lithium values for the six stars with the highest lithium abundance in their respective metallicity bins. The red circles show the results from the work by \citet{Guiglion19}.
}
\label{fig:ALiFEHVsini}
\end{figure}

To study the Li depletion and the Li abundance in the ISM one has to look at the most Li abundant stars which reflect the initial Li abundances in the ISM (e.g., \citealt{LambertReddy2004}, \citealt{DelgadoMena15}, \citealt{Guiglion16,Guiglion19}). 
Following similar analyses, we binned our data every 0.15~dex in [Fe/H], except for the boundary regions as there are too few stars, and we made the metallicity bins of 0.3~dex for both the most metal poor ($-0.80$ to $-0.50$~dex) and the metal rich side (from +0.25 to +0.55~dex). In order  to be consistent with previously mentioned studies, we took the six most lithium rich stars in every bin. The error bars represent the standard deviation between those six stars. 
The relation is shown in Figure~\ref{fig:ALiFEHVsini}. 
When we try to trace the lithium abundance in the environment from which stars were formed, one has to keep in mind that lithium is easily destroyed in stellar interiors and on the other hand,
some lithium can be produced as well. Thus, only dwarf stars with the largest Li abundances should be considered, as such stars have only burned lithium in their photospheres and do not contain any that would be freshly synthesized. 
In Figure~\ref{fig:ALiFEHVsini}, we see that considering only the largest values of the Li abundances in dwarf stars, we confirm that there is a decrease of lithium at supersolar metallicities which cannot be explained by the most recent models  \citep[see][and discussions therein]{Romano01,Prantzos17,Fu18}. This work shows that the closest and brightest stars in a  small volume of space in the solar neighborhood also exhibit this trend. The largest lithium abundance of 3.0~dex in our study is reached at the solar metallicity and drops by 0.69~dex in the range of [Fe/H] from  +0.10 to +0.55~dex. Furthermore, stars in Figure~\ref{fig:ALiFEHVsini} are color-coded by {\it{v}}\,sin\,{\it{i}} and we find some evidence that \ALi~seems to increase with increasing stellar rotation speed. 

\subsection{Carbon and Oxygen Abundances in the Thin and Thick Galactic Disk}

We determined carbon and oxygen abundances for 78 observed stars including the Sun. The forbidden oxygen [O\,{\sc i}] line at 6300~\AA~was severely blended by telluric lines in quit many spectra of our sample of stars. This was the main reason why the total number of stars with determined carbon and oxygen abundances was lower than that for lithium. Stars in our sample were divided into the thin and thick disks according their chemical composition (see \citealt{Mikolaitis19} for more details).

Figures~\ref{fig:CFe_vs_FeH} and \ref{fig:OFe_vs_FeH} show the carbon and oxygen abundances versus metallicity. The thin-disk stars are marked as circles, the squares show the thick-disk stars. All the stars are color-coded by the C/O number ratio (note that C/O $\neq$ [C/O]). As expected, both [C/Fe] and [O/Fe] decrease with time due to the increasing production of iron by SNe~Ia supernovae. The carbon clearly follows iron more closely than oxygen. 
The data obtained in our work are in agreement with previous studies \citep[e.g.,][]{Bensby06,Brewer17,Luck17}. 
In theory, oxygen is synthesized mostly in massive stars, while carbon is produced in stars of all masses. Thus, the ejection of some carbon is delayed in time with respect to oxygen, so the [C/O] offers a relative age clock for stellar systems \citep{Tinsley79}.

\textbf{\begin{figure}
\resizebox{\hsize}{!}{
\includegraphics[width=\columnwidth]{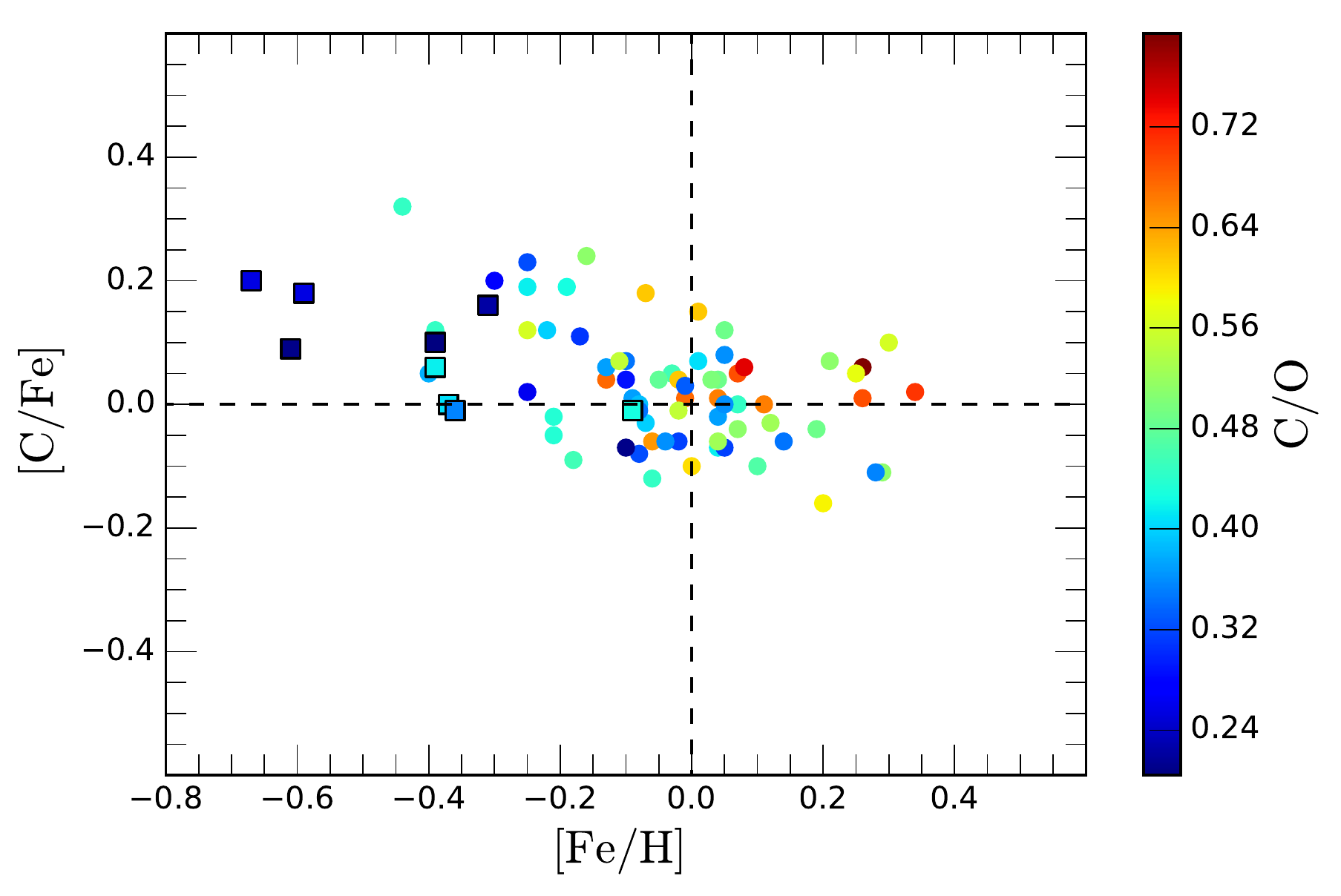}
}
\caption{[C/Fe] versus [Fe/H]. Stars are color-coded by their C/O number ratio. The thin-disk stars are shown as circles, the thick-disk stars as squares. }
\label{fig:CFe_vs_FeH}
\end{figure}}

\textbf{\begin{figure}
\resizebox{\hsize}{!}{
\includegraphics[width=\columnwidth]{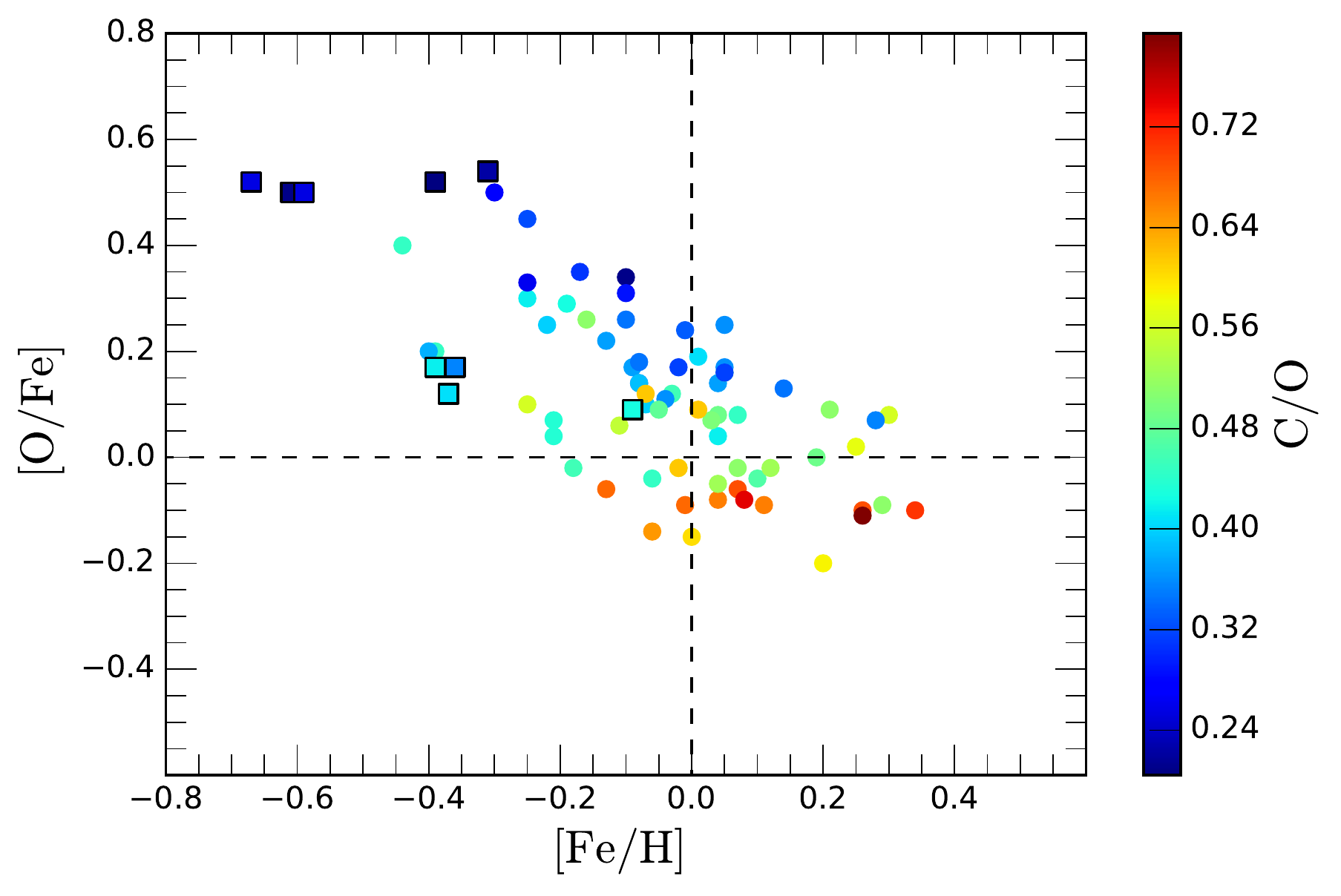}
}
\caption{[O/Fe] versus [Fe/H]. The meaning of symbols are the same as in Figure~\ref{fig:CFe_vs_FeH}. }
\label{fig:OFe_vs_FeH}
\end{figure}}

\textbf{\begin{figure*}
\resizebox{\hsize}{!}{
\includegraphics[width=\columnwidth]{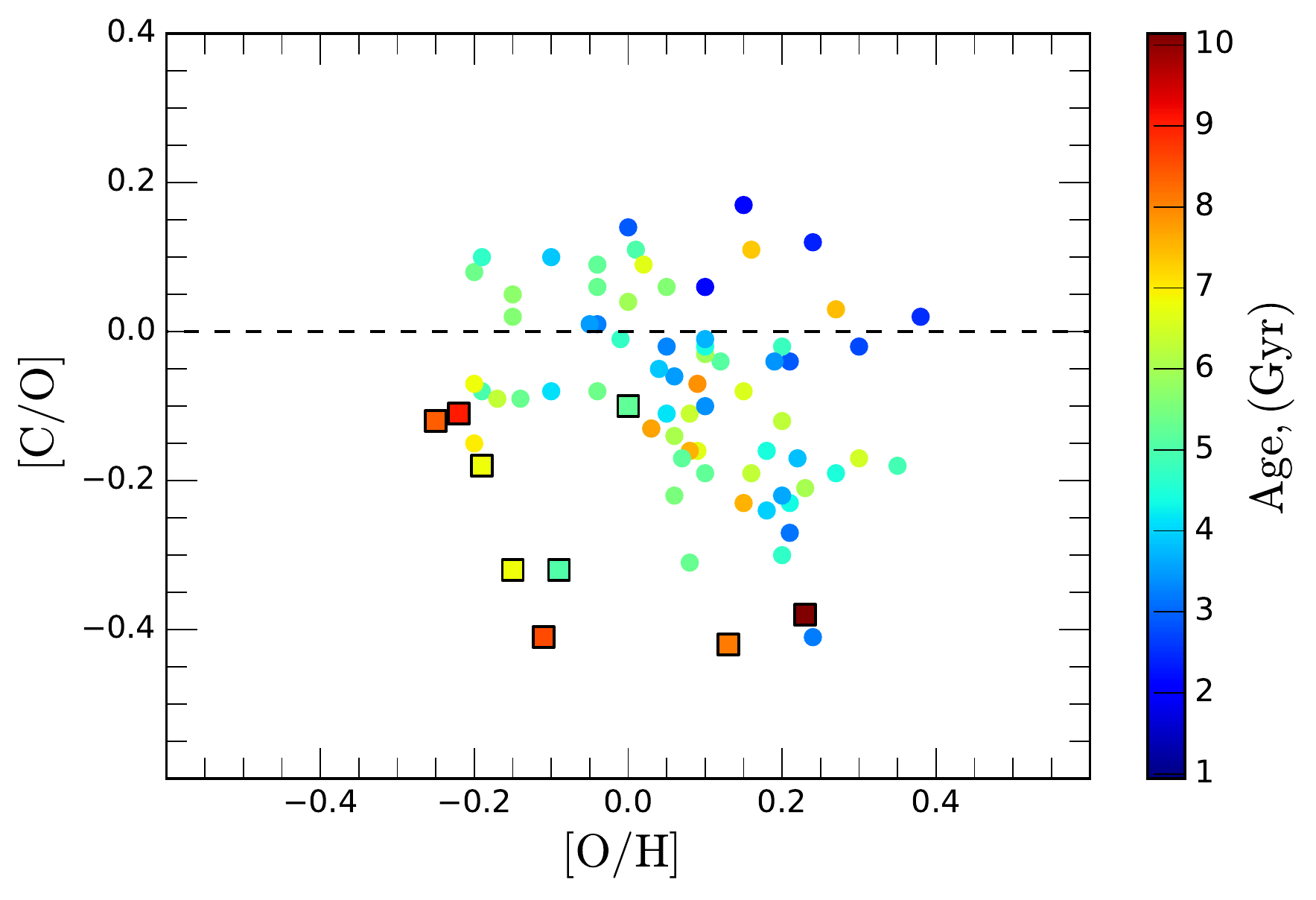}
\includegraphics[width=\columnwidth]{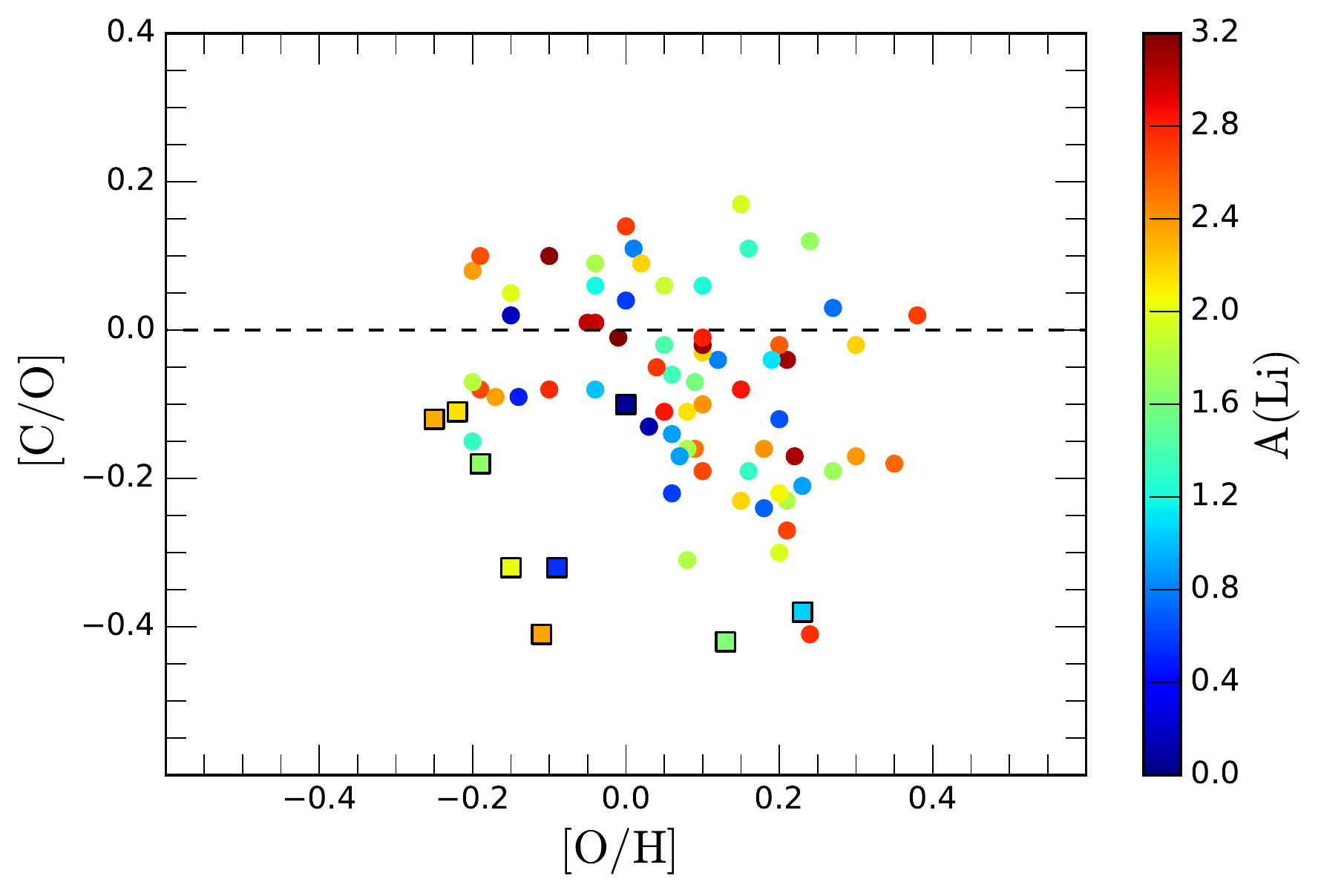} 
}
\caption{[C/O] versus [O/H]. Left: our results for thin-disk stars are shown as circles, thick-disk as squares. The stars are color-coded by age. Right: The same plot with stars color-coded by their lithium abundances.}
\label{fig:CO_OH_CO_compa}
\end{figure*}}

Knowing the fact that oxygen is mostly produced in massive stars on a relatively short timescale (see \citealt{Cescutti09}), [C/O] as a function of [O/H] show results of the carbon evolution.
The change in carbon-to-oxygen as a function of oxygen-to-hydrogen strongly depends on the yields and timescales of carbon production in various types of stars. 
In Figure~\ref{fig:CO_OH_CO_compa} we see a systematic difference between the thin and thick disks in the [C/O] versus [O/H] plane. The thick-disk stars lie on a different [C/O] sequence compared to the thin-disk stars which is shifted by $\sim 0.3$~dex to lower [C/O] values. To explain this, carbon produced in both low-mass and massive stars has to be included.

The difference between thin- and thick-disk stars in the carbon-to-oxygen abundance ratio as a function of oxygen-to-hydrogen abundance ratio plane for dwarf stars has previously been demonstrated by \citet{Bensby06} using 35 thin- and 16 thick-disk stars and by \citet{Nissen14} who investigated 57 thin- and 25 thick-disk stars. \citet{Bensby06} used the forbidden carbon ([C\,{\sc i}] at 8727 \AA) and oxygen ([O\,{\sc i}] at 6300~\AA) lines. \citet{Nissen14} investigated the high-excitation permitted C\,{\sc i} lines at 5052 and 5380~\AA, and O\,{\sc i} at 7774~\AA). The stars in the work by \citet{Bensby06} were assigned to populations based on statistical probability of their kinematics.  \citet{Nissen14} derived C and O from two data samples obtained using different spectrographs: HARPS, FEROS, UVES, and FIES, and stars were assigned to the separate disks by both the chemical and kinematical approach. Our results confirm the separation of thin- and thick-disk stars in the [C/O] versus [O/H] plane using the homogeneous analysis of C$_2$ bands at 5135~\AA~and 5635~\AA~and the [O\,{\sc i}] forbidden line at 6300~\AA, and taking into account the molecular equilibrium of these elements.

In addition, we see that both the thin- and thick-disk dwarfs have no [C/O] correlation with \ALi~in the [C/O] versus [O/H] plane (the right panel in Figure~\ref{fig:CO_OH_CO_compa}). To confirm this, we also checked the recent data by \citet{Luck17} and found the same. 

\subsection{Li, C, and O in Stars with Planets}

\begin{figure}
	\graphicspath{ {} }
	\includegraphics[width=\columnwidth]{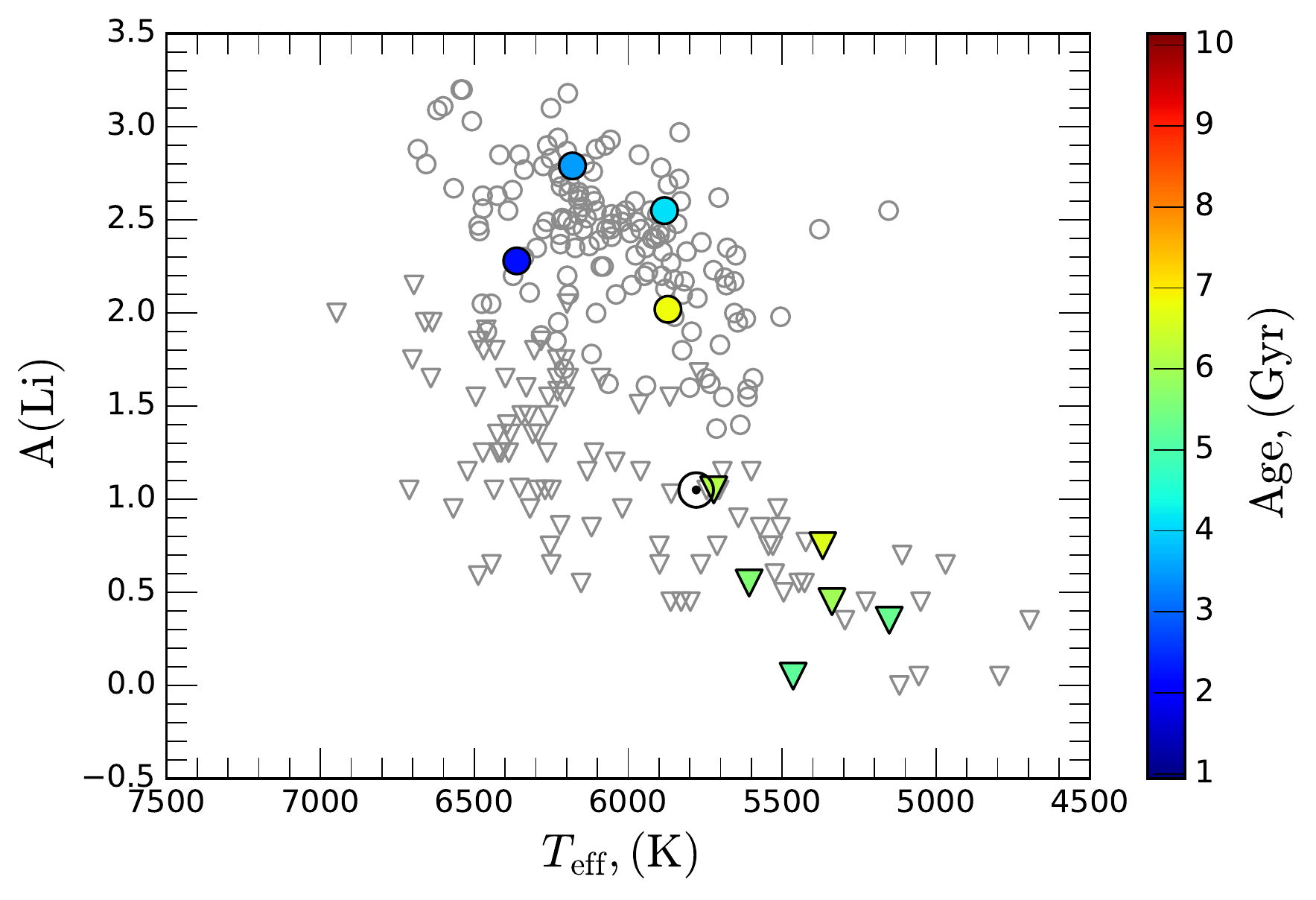}
    \caption{Lithium abundance as a function of the effective temperature. The stars with so far detected planets with measurable \ALi~and upper limits are shown as circles or triangles, respectively, and color-coded by their age values.}
    \label{fig:Li_planets}
\end{figure}

Several previous studies have shown that stars with detected planets have different abundance patterns  compared to those with so far no planets detected. The results, however, are still debatable \citep[see, e.g.,][and references therein]{Israelian09,  Tucci14, DelgadoMena14, daSilva15, Mishenina16, Suarez17, Bensby18, Luck18b, Pavlenko18, Carlos19}. 

We found that 10 stars in our sample have exoplanets detected. In Table~\ref{table:exoplanets} we collected data on the exoplanets identified around those stars \citep{Akeson13}. For 5 out of 10 we have both carbon and oxygen abundance measurements and all 10 have lithium abundances determined. For convenience, in Table~\ref{table:exoplanets} we also present the determined Li abundances and C/O number ratios from our work as well as the Mg/Si number ratios from \cite{Mikolaitis19}, obtained from the same observational data and stellar atmospheric parameters as in our work. 

In Figure~\ref{fig:Li_planets}, we show lithium abundances for the sample stars and mark with colored symbols the stars with confirmed exoplanets. Four out of 10 planet-hosting stars have confidently measured \ALi~values. 
Four stars with relatively high effective temperatures ($<$\Teff$>=6070\pm200$~K) with confirmed exoplanets have an average $<A{\rm(Li)}>\sim2.4$.  
The remaining six stars have much lower \ALi~abundances ($<A{\rm(Li)}>=0.50\pm0.3$) with only upper limits measured and on average have lower effective temperatures ($<$\Teff$>=5440\pm185$~K). 

\begin{figure}
	\graphicspath{ {} }
	\includegraphics[width=\columnwidth]{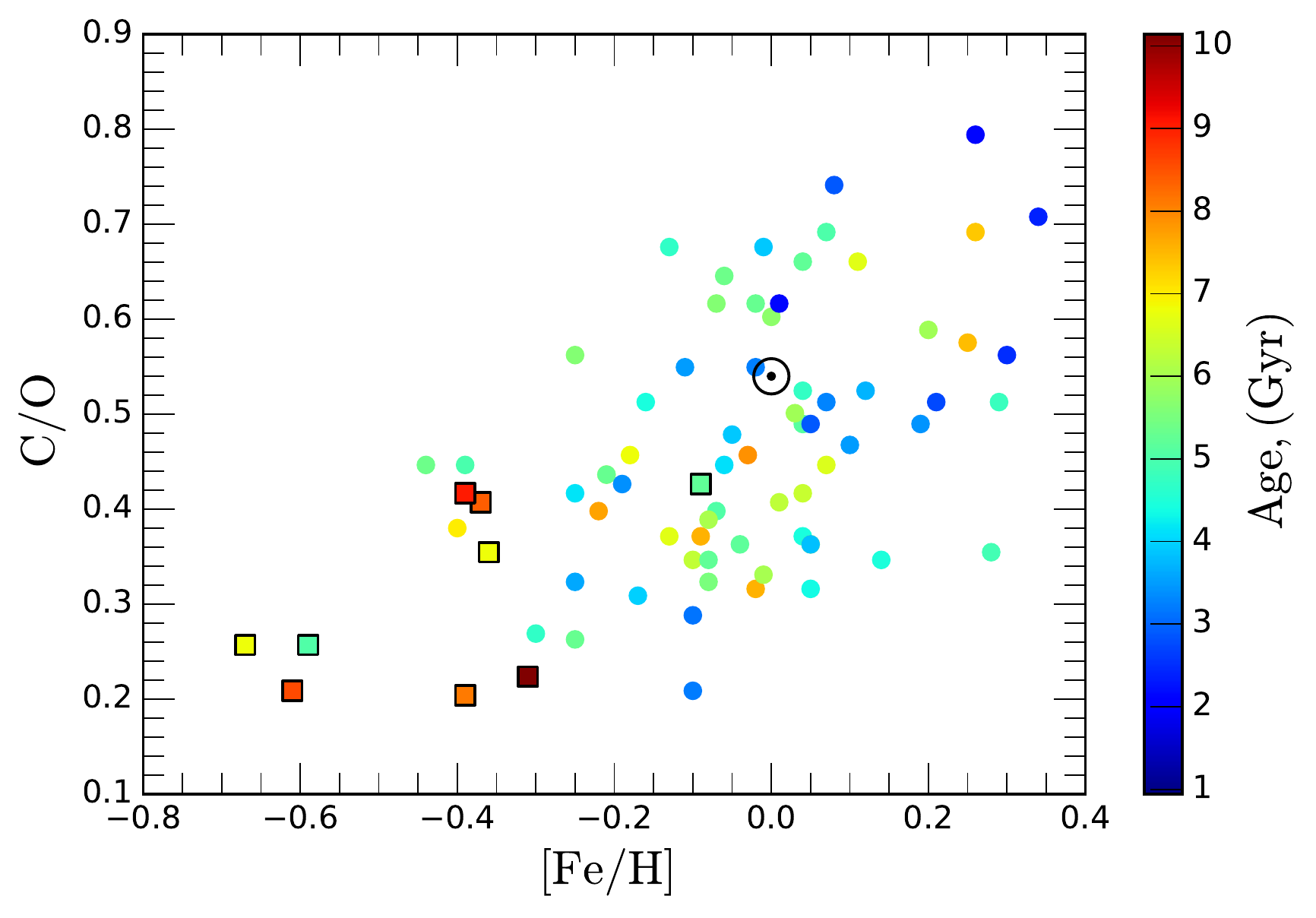}
    \caption{C/O number ratio as a function of metallicity. Stars are color-coded by age. The thin-disk stars are shown as circles, the thick-disk stars -- as squares. }
    \label{fig:CO_vs_Age}
\end{figure}

From the available data, we cannot see a discernible difference between the planet hosts and the stars with no planets detected regarding their lithium content (Figure~\ref{fig:Li_planets}). From a larger sample of F, G, and K dwarfs of the Southern Hemisphere containing 100 planet hosts, \citet{Luck18b} also came to the same conclusion. Moreover, \citet{Carlos19} emphasize a strong connection between lithium depletion and stellar ages, and notice that there is no significant difference in lithium
depletion between known planet-host stars and stars with
no planets detected, when the lithium abundance
and stellar age correlation is analyzed. However, there could be planet-hosting stars like our Sun with quite low lithium abundances. 

Theoretical studies suggest that the C/O number ratio as well as Mg/Si are important in determining the mineralogy of extraterrestrial planets. While the C/O ratio controls the amount of carbides and silicates formed in planets, the Mg/Si can tell us about the silicate mineralogy \citep[e.g.,][]{Bond10,Madhusudhan14,Piso16}. 
For stars with Mg/Si\,$<1$, terrestrial planets will have a magnesium depleted mineralogy different from that of the Earth. For stars with ${\rm Mg/Si}\sim 1\div 2$, Mg is distributed between olivine and pyroxene, leading to rocky planet compositions similar to the Earth. In case of Mg/Si\,$>2$, all available Si is consumed to form olivine with excess Mg available to bond with other elements as MgO or MgS \citep{Bond10}.
Furthermore, the C/O number ratio as well as Mg/Si are important in the formation, atmospheric chemistry, interior structure for all types of exoplanets, and for the plate tectonics and habitability of terrestrial planets. For example, in order to constrain the interior structure of rocky exoplanets the stellar elemental abundances (such as Fe, Si, and Mg) are key constraints to reduce degeneracy in interior structure models and to constrain the mantle composition \citep{Dorn15}.
Recent observational data showed that C/O varies among planetary atmospheres and host stars, and even among planets within the same planetary system \citep{Brewer17, Suarez18}.

\begin{figure}
\resizebox{\hsize}{!}{
\includegraphics[width=0.7\columnwidth]{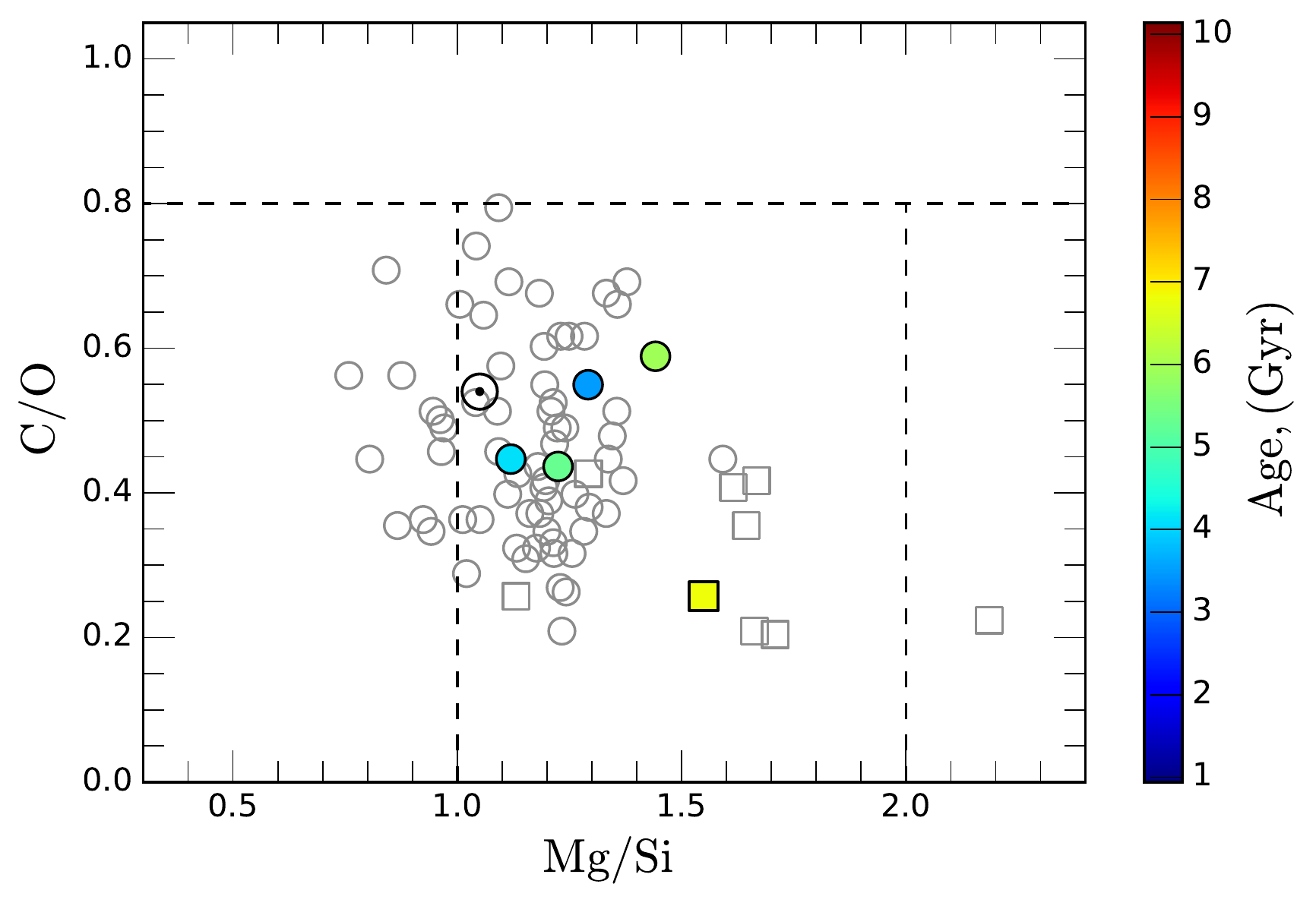} 
}
\caption{C/O number ratio as a function Mg/Si number ratio. Our results for the thin-disk stars are shown as circles, the thick-disk -- squares. Mg/Si number ratios are taken from \cite{Mikolaitis19}. Stars with so far detected planets are color-coded by age. The dashed lines divide stellar parameter spaces were gaseous (if Mg/Si~$<1$) or rocky ($1<{\rm Mg/Si}<2$) exoplanets may form (see \citealt{Suarez18} and the text for more explanations). }
\label{fig:CO_MgSi_compa}
\end{figure}

\begin{figure}
\resizebox{\hsize}{!}{
\includegraphics[width=0.73\columnwidth]{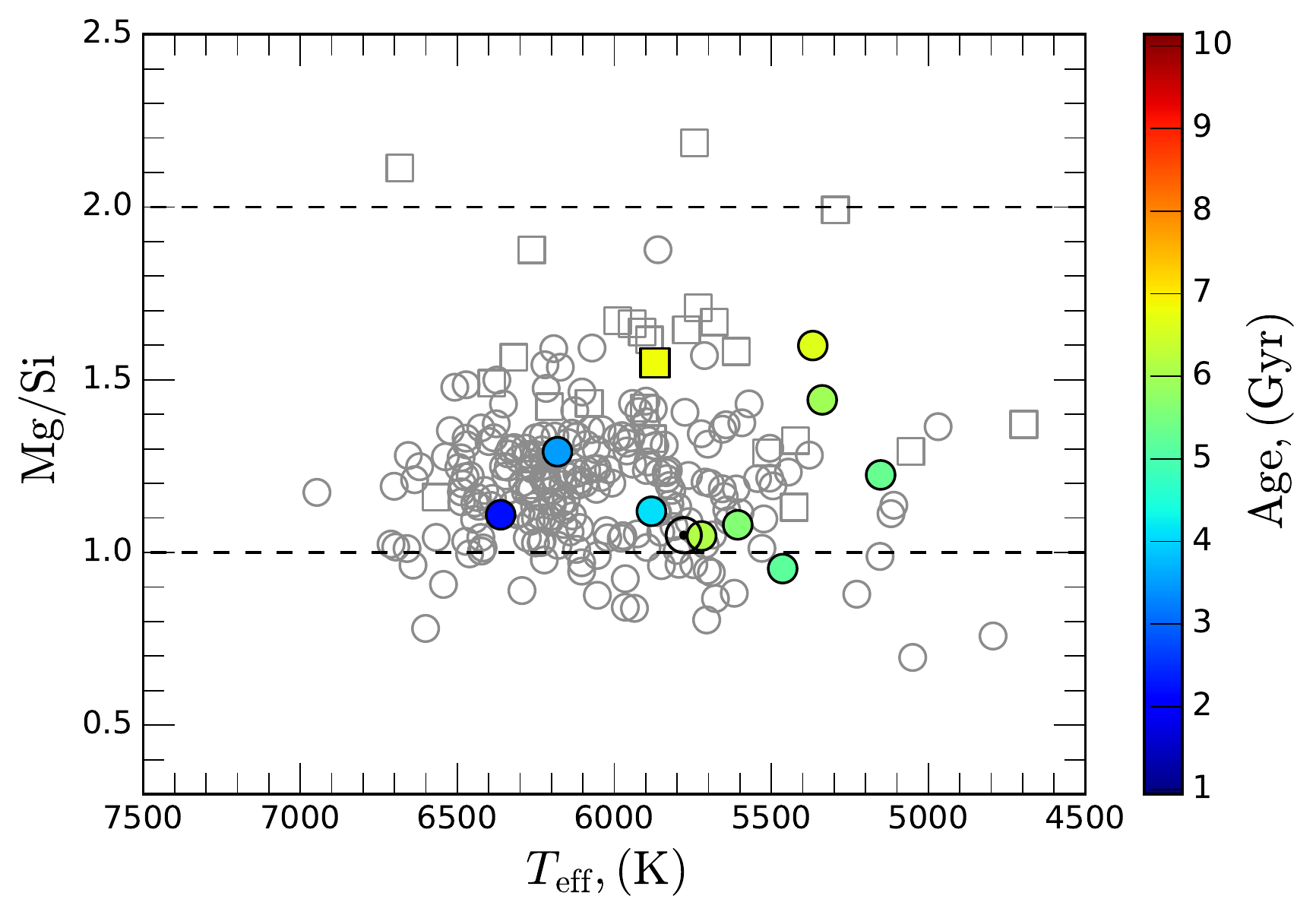}
}
\caption{Mg/Si number ratio as a function of the effective temperature.  All 10 stars with so far detected planets are color-coded by age. 
}
\label{fig:MgSi_Teff_compa}
\end{figure}

Figure~\ref{fig:CO_vs_Age} shows the C/O number ratio as a function of metallicity. Stars are color-coded by age. 
First of all, we notice the nine stars with predominantly supersolar metallicities that have the C/O number ratios larger than 0.65, thus according to e.g., \citet{Moriarty14} they may form carbon-rich rocky planets.  This makes an unexpectedly large percentage, 12\,\% of the sample stars, which is a hundred times larger than it was found by \citet{Brewer16b}. The percentage of 0.13\,\% for the carbon-enhanced dwarfs in the solar neighborhood predicted by  \citet{Brewer16b} is indeed probably underestimated since, e.g., \citet{Suarez18} found about 1\,\% of their sample stars having C/O ratio even larger than 0.8. Our stellar sample does not contain stars with C/O$> 0.8$, which necessarily should have carbon-rich rocky planets if detected. 
The mean C/O ratio for 53 sample stars within $\pm0.2$~dex of the solar metallicity is $0.47\pm0.12$, this value falls exactly to the C/O peak found for the solar-vicinity stars by \citet{Brewer16b} and \citet{Suarez18}.   

\begin{table*}
\renewcommand{\tabcolsep}{0.95mm}
 \caption{Stars of Our Sample with Confirmed Eexoplanets.}
 \label{tab:exoplanets}
 \begin{tabular}{ccccclcc}
  \hline
  \hline
  { \raisebox{-1.5ex}[0cm][0cm]{TYC ID}} & { \raisebox{-1.5ex}[0cm][0cm]{Planet}} & Planet mass &Semi-major axis & Orbital period & & Host stars$^a$\\
 & &($M_{\text{Jup}}$)& (au) & (day) & \ALi & C/O & Mg/Si\\
  \hline
  
      \noalign{\smallskip}
 2099-2717-1 & HD 164922 b & $0.31\pm{0.05}$ & $2.10\pm{0.04}$ & $1155\pm{23}$  & 0.45$^{b}$  & 0.59 &  1.44 \\
  & HD 164922 c & $0.04\pm{0.005}$ & $0.34\pm{0.0015}$ & $75.77_{-0.056}^{\text{+}0.058}$  \\
  
   \noalign{\smallskip}
 2103-1620-1 & HD 164595 b & $0.05\pm{0.00856}$ & $0.23$ & $40\pm{0.24}$  & 1.05$^{b}$  & ... & 1.05 \\

   \noalign{\smallskip}
 2595-1464-1 & HD 155358 b & $0.99\pm{0.08}$ & $0.63\pm{0.02}$ & $194.3\pm{0.3}$   & 2.02 & 0.26 & 1.55 \\
 & HD 155358 c & $0.82\pm{0.07}$ & $1.02\pm{0.02}$ & $391.9\pm{1}$  \\
 
   \noalign{\smallskip}
 2648-2151-1 & HD 178911 B b & $8.03\pm{2.51}$ & $0.34\pm{0.01}$ & $71.48\pm{0.02}$ & 0.55$^{b}$  & ... & 1.08 \\
 
  \noalign{\smallskip}
3067-576-1 & 14 Her b & $4.66\pm{0.15}$ & $2.93\pm{0.08}$ & $1773.4\pm{2.5}$  & 0.75$^{b}$  & ... & 1.60 \\

 \noalign{\smallskip}
 3501-1373-1 & HD 154345 b & $0.82\pm{0.07}$ & $4.21\pm{0.10}$ & $3341.56\pm{93}$ & 0.05$^{b}$  & ... & 0.95 \\

\noalign{\smallskip} 
4436-1424-1 & $\psi^{1}$ Draconis B b & $1.53\pm{0.10}$ & $4.43\pm{0.04}$ & $3117\pm{42}$  & 2.79 & 0.55 & 1.29 \\

\noalign{\smallskip} 
4494-1346-1 & HD 7924 b & $0.02\pm{0.00}$ & $0.06\pm{0.00}$ & $5.4\pm{0.00025}$  & 0.35$^{b}$ & 0.44 & 1.23 \\

 \noalign{\smallskip}
              & HD 7924 c
 & $0.02_{-0.0022}^{\text{+}0.0023}$ & $0.11_{-0.0014}^{\text{+}0.0013}$ & $15.3_{-0.0033}^{\text{+}0.0032}$  \\
 \noalign{\smallskip}
              & HD 7924 d
 & $0.02_{-0.0025}^{\text{+}0.0025}$ & $0.16_{-0.0019}^{\text{+}0.0018}$ & $24.45_{-0.017}^{\text{+}0.015}$  \\
 
 \noalign{\smallskip}
 4532-2096-1 & HD 33564 b & $9.1$ & $1.1$ & $388\pm{3}$  & 2.28 & ... & 1.11 \\

 \noalign{\smallskip}
 4575-1336-1 & HD 150706 b & $2.71_{-0.66}^{\text{+}1.14}$ & $6.7_{-1.4}^{\text{+}4.0}$ & $5894_{-1498}^{\text{+}5584}$ & 2.55 & 0.45 & 1.12\\
 
  \hline
 \end{tabular}
 \flushleft
       {\bf{Notes.}} Data were taken from  NASA Exoplanet Archive \citep{Akeson13} on 2019 July 31. \\
       $^a$ Our study results.\\
       $^b$ Upper limits.\\

 \label{table:exoplanets}
\end{table*}

In Figure~\ref{fig:CO_MgSi_compa} we show C/O number ratios as a function of Mg/Si. Stars with detected planets are color-coded by age. The Sun with our determined values of C/O$_{\sun} = 0.54$ and Mg/Si$_{\sun} = 1.05$ is marked as well.
About 89\,\% of stars have Mg/Si values ranging from 1.0 to $\sim 2.0$ with the mean value of Mg/Si=$1.26 \pm 0.17$, which would lead to rocky planet compositions close to that of the Earth. The remaining $\sim$10\,\% of the stars have Mg/Si~$<1.0$, where terrestrial planets would have a magnesium depleted mineralogy with the mean C/O~$\simeq 0.48 \pm 0.10$. The thin-disk stars have C/O ratio in the range from 0.2 to 0.8 with the mean Mg/Si=$1.16 \pm 0.16$. Whereas, the majority of the thick-disk stars (on average older stars) have C/O ratios between 0.2 and 0.4 with the mean Mg/Si=$1.61 \pm 0.28$. 
The Mg/Si number ratios of all 249 stars are shown in  Figure~\ref{fig:MgSi_Teff_compa} as a function of the effective temperatures. Several thick-disk stars have Mg/Si$>2.0$. 

The percentage of stars with exoplanets in our sample so far is very small, but with the new data from the NASA \textit{TESS} space mission, there is no doubt, our homogeneous high-resolution abundance analysis in planet hosts will present a useful case for future studies.

\section{Summary and conclusions}\label{sec:Discussion and summary}

In the presented work, we homogeneously investigated abundances of lithium, carbon, and oxygen in a sample of 249 bright dwarf stars of the northern sky with a rather broad metallicity range.
The investigated stars have metallicities, effective temperatures, and ages between \mbox{(--0.7 $\div$ 0.4) dex}, \mbox{(5000 $\div$ 6900) K}, and \mbox{(1 $\div$ 12) Gyr}, accordingly.
We determined lithium abundances for all investigated stars and carbon and oxygen abundances for 78 stars of the sample. 
The stars belong mainly to the thin Galactic disk. 

\ALi~is predominantly higher in the hottest, youngest, and more massive stars of our sample as expected. The decrease of \ALi per 1000~K equals to $0.4 \pm 0.1$~dex for stars with  \ALi\,$\geq$~2.0. The lithium abundance correlation with the effective temperature is similar for both the lithium-rich stars with \ALi\,$\geq$~2 and for those with \ALi$<2$. 

We identified dwarf stars with similar ages, atmospheric parameters, and rotational velocities, but significantly different \ALi values. We speculate that in addition to the above-mentioned main parameters, lowered lithium abundances in these solar-vicinity dwarfs could be caused by external factors such as the Galactic kinematic evolution and partially by presence of planets.

We confirm that chromospherically active dwarfs may have quite high lithium abundances. We found two such stars in our investigated sample. 

Our work shows that even in a small volume of space in the solar neighborhood stars exhibit decrease of lithium at supersolar metallicities -- the phenomenon that cannot be explained even by the most recent models. The highest lithium abundance of 3.00~dex in our study is reached at the solar metallicity and drops by 0.7~dex in the [Fe/H] range from +0.10~to~+0.55~dex.

In the investigated sample of stars, we cannot see a discernible difference between the planet hosts and the stars with no planets detected regarding their lithium content.
 Both the thin- and thick-disk dwarfs have no [C/O] correlation with A(Li). 

Nine stars with predominantly supersolar metallicities, i.e. about 12\,\% of the sample of 78 stars with C and O abundances determined, have the C/O number ratios larger than 0.65, thus may form carbon-rich rocky planets. These stars could be interesting targets for space (e.g., \textit{TESS} and \textit{PLATO}) and ground-based  exoplanet search surveys. About 89\,\% of our sample stars have the mean value of Mg/Si~$= 1.26 \pm 0.17$, i.e. Mg/Si ratios are in the range between one and two, which would lead to rocky planet compositions close to that of the Earth. The remaining $\sim$10\,\% of the stars have Mg/Si$<$1.0, where terrestrial planets would have a magnesium depleted mineralogy with the mean C/O~$\simeq 0.48 \pm 0.10$.

The percentage of stars with confirmed exoplanets in our sample is still small and we plan to contribute in the analysis of the key elements like lithium, carbon, and oxygen in planet hosts with the homogeneous high-resolution data at once the new data from the NASA \textit{TESS} space mission will be available for stars in the Northern Hemisphere. 

\section*{Acknowledgements}

We gratefully acknowledge the grant from the European Social Fund via the Lithuanian Science Council (LMTLT) grant No. 09.3.3-LMT-K-712-01-0103. We thank the anonymous referee for the prompt and valuable report. This research has made use of the NASA Exoplanet Archive, which is operated by the California Institute of Technology, under contract with the National Aeronautics and Space Administration under the Exoplanet Exploration Program.


\begin{appendix}\label{appendix}
\section{Information about the online table}

Information about columns and meanings of acronyms used in presenting the full online table of results for the investigated 249 stars. The main atmospheric parameters (\Teff, \logg, \FeH, \vt) of stars were taken from \citet{Mikolaitis18, Mikolaitis19}.

\begin{longtable}{ c l l l }
\hline
\hline
\noalign{\smallskip}
  Column  & Label & Unit & Comment \\
\noalign{\smallskip}
\hline
\noalign{\smallskip}
(1)  &   Name   		&           & Star name TYC      \\
(2)  &   \ALi		    &    dex    & \ALi~abundance      \\
(3)  &   \textit{M}$|$\textit{U}            &           &  M--measurable, U--upper limit of \ALi    \\
(4)  &   \Vsini		    &	 \kms	&  Rotational velocity   \\
(5)  &   [C/H]		    &     dex   &  Carbon-to-hydrogen ratio       \\
(6)  &\textit{e}\_[C/H]	        & 	dex	    &   Error on  carbon-to-hydrogen ratio       \\
(7)  &$\rm n{_C}$       & 	        &   Number of lines    \\
(8)  &   [O/H]		    &     dex   &   Oxygen-to-hydrogen ratio       \\
(9)  &\textit{e}\_[O/H]	        & 	dex	    &   Error on oxygen-to-hydrogen ratio       \\
(10)   &$\rm n{_O}$     & 	        &   Number of lines    \\
(11)  &  C/O		    &        &   Carbon-to-oxygen number ratio    \\
(12)  &Thin$|$Thick        & & Thin-disk star ``0''$|$Thick-disk ``1''\\
(13) & Planet        & & Have any planet ``1''$|$Have not ``0''\\
(14) &\Teff             & K         & Effective temperature\\
(15) &\textit{e}\_\Teff           &K          & Error on effective temperature\\
(16) &\logg              &  dex   & Surface gravity\\
(17) &\textit{e}\_\logg         &  dex   & Error on surface gravity\\
(18) &[Fe/H]             &  dex   & Metallicity \\
(19) &\textit{e}\_[Fe/H]          &  dex   & Error on metallicity \\
(20) &\textit{V}t                  &   \kms  & Microturbulence velocity\\
(21) &e\_\textit{V}t                &  \kms   & Error on microturbulence velocity\\
\noalign{\smallskip}
\hline
    \label{tab:results_log}
\end{longtable}

\end{appendix}





\bibliography{Stonkute20}{}
\bibliographystyle{aasjournal}



\label{lastpage}






\end{document}